\documentclass[journal]{IEEEtran}

\ifCLASSINFOpdf

\else

\fi

\usepackage{graphics}
\usepackage{epsfig}
\usepackage{mathptmx}
\usepackage{times}
\usepackage{amsmath}
\usepackage{amssymb}
\usepackage{eurosym}
\usepackage{textcomp}
\usepackage{graphicx,subfigure}
\usepackage{fancybox}

\usepackage{longtable}
\usepackage{rotating}
\usepackage{array}
\usepackage{multicol}
\usepackage{multirow}

\newtheorem{remark}{Remark}

\newtheorem{proposition}{Proposition}

\begin{document}

\title{An efficient model-free setting for longitudinal and lateral vehicle control. Validation through the interconnected pro-SiVIC/RTMaps prototyping platform}

\author{Lghani MENHOUR, Brigitte d'ANDR\'EA-NOVEL, Michel FLIESS, Dominique GRUYER \\ and Hugues MOUNIER
\thanks{L. Menhour is with Universit\'e de Reims, 9, Rue du Qu\'ebec, 10000 Troyes, France {\tt\small lghani.menhour@univ-reims.fr}}
\thanks{B. d'Andr\'ea-Novel is with Centre de Robotique, Mines ParisTech, PSL Research University, 60 boulevard Saint-Michel, 75272 Paris cedex 06, France. {\tt\small brigitte.dandrea-novel@mines-paristech.fr}}
\thanks{M. Fliess is with LIX (CNRS, UMR 7161), \'Ecole polytechnique, 91128 Palaiseau, France. {\tt\small Michel.Fliess@polytechnique.edu} and AL.I.E.N. (ALg\`{e}bre pour Identification et Estimation Num\'{e}riques), 24-30 rue Lionnois, BP 60120, 54003 Nancy, France. {\tt\small  michel.fliess@alien-sas.com}}
\thanks{D. Gruyer is with IFSTTAR-CoSys-LIVIC, 77 rue des Chantiers, 78000 Versailles, France. {\tt\small dominique.gruyer@ifsttar.fr}}
\thanks{H. Mounier is with L2S (UMR 8506), CNRS -- Sup\'elec -- Universit\'{e} Paris-Sud, 3 rue Joliot-Curie, 91192 Gif-sur-Yvette, France. {\tt\small hugues.mounier@lss.supelec.fr}
}}


\markboth{\tt IEEE Transactions on Intelligent Transportation Systems}%
{Shell \MakeLowercase{\textit{et al.}}: Bare Demo of IEEEtran.cls for IEEE Journals}

\maketitle

\begin{abstract}
In this paper, the problem of tracking desired longitudinal and lateral motions for a vehicle is addressed. Let us point out that a ``good''  modeling is often quite difficult or even impossible to obtain. It is due for example to parametric uncertainties, for the vehicle mass, inertia or for the interaction forces between the wheels and the road pavement. To overcome this type of difficulties, we consider a model-free control approach leading to ``intelligent'' controllers. The longitudinal and the lateral motions, on one hand, and the driving/braking torques and the steering wheel angle, on the other hand, are respectively the output and the input variables. An important part of this work is dedicated to present simulation results with actual data. Actual data, used in Matlab as reference trajectories, have been previously recorded with an instrumented Peugeot 406 experimental car. The simulation results show the efficiency of our approach. Some comparisons with a nonlinear flatness-based control in one hand, and with a classical PID control in another hand confirm this analysis. Other virtual data have been generated through the interconnected platform SiVIC/RTMaps, which is a virtual simulation platform for prototyping and validation of advanced driving assistance systems.
\begin{center}
\textbf{Keywords}
\end{center}
Longitudinal and lateral vehicle control, model-free control, intelligent controller, algebraic estimation, flatness property, classical PID controllers, ADAS (Advanced Driving Assistance Systems)
\end{abstract}

\section{Introduction}
\label{introd}

The vehicle longitudinal and lateral control problem has been widely investigated in the literature via model-based techniques (see, \textit{e.g.}, \cite{Ackermann95, Novel01, Choi09, Cerone09, Chou05, Fernandez11, formentin, Fuchsumer2005,Hatipoglu03, Khodayari10,Manceur13, Menhour13a, Ge15, Wei15, Nobe01, Odenthal99, Poussot11a, Tagne16, Villagra09}, and the references therein). Their performances are guaranteed in the vicinity of the model used for their implementation.

\begin{figure}[!ht]
\centering
\includegraphics[scale=0.5]{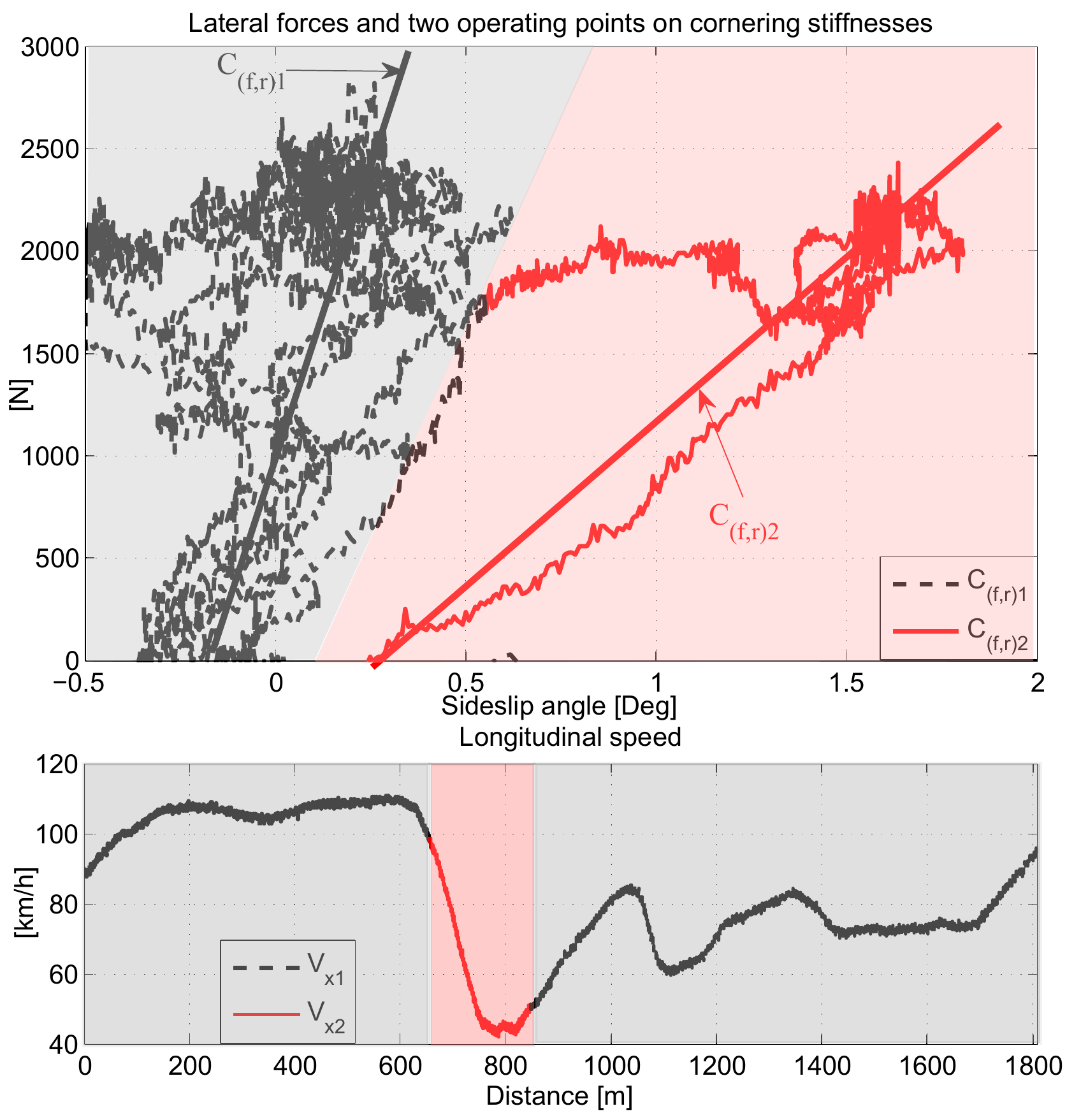}
\caption{Experimental lateral tire force characteristic and its nonlinear behavior}
\label{switching_rule}
\end{figure}

In this paper, a description of the evolution of our developments on longitudinal and lateral vehicle control problem is given. 
Thereafter, the poor knowledge of vehicle mathematical models has led us to develop a vehicle controller based on \emph{model-free} setting \cite{ijc13}. In fact, obtaining of a ``good'' mathematical modeling is a difficult task, if not an impossible one, since uncertainties and disturbances, like frictions and tire nonlinear behaviors, should be taken into account. The vehicle behavior is highly dependent on tire road forces which are usually modeled by linear tire models. Fig. \ref{switching_rule} illustrates an example of the tire characteristic obtained during an experimental braking maneuver on a real race track. This maneuver has highlighted the different nonlinear dynamics of the tire forces which cannot be characterized by a linear tire model. Unfortunately, there are other emergency driving situations for which simplified vehicle models cannot provide a realistic behavior of the actual car, such as the rollover phenomenon due to high values of Load Transfer Ratio (LTR), and the under-steering and over-steering behaviors due to high values of front or rear sideslip angles.

The parametric dependency and the tedious features vanish here thanks to a new model-free approach where:
 
\begin{itemize}
\item the need to exploit the flatness property of a simplified model disappears,
\item the flat output depending on uncertain parameters is replaced by a simpler and perhaps more natural one, which is the lateral deviation.
\end{itemize}

Our paper is organized as follows. A brief description of the required vehicle models used for simulation or control purposes is given in Section \ref{Section_1}. A short summary on the model-free control approach with application to the design of a longitudinal/lateral vehicle control and a lateral deviation computation algorithm are presented in Section \ref{Section_3}. Section \ref{Section_4} displays several numerical simulations established using actual data as reference trajectories and a full nonlinear 10DoF vehicle model of an instrumented Peugeot 406 experimental car.
The obtained results show the relevance and efficiency of our approach. Some comparisons are provided with a nonlinear flatness-based controller and with classical PID controllers, briefly recalled in Appendix \ref{appendix1}, in order to highlight the advantages of model-free control. Secondly, a validation of model-free control is conducted under an interconnected platform SiVIC/RTMaps (See \cite{Gruyer10a}). This platform provides an efficient  simulation tool in order to replace an actual system (sensors, actuators, embedded algorithms) by a virtual one in a physically realistic  environment. It allows advanced prototyping and validation of perception and control algorithms including quite realistic models of vehicles, sensors and environment. Some concluding remarks are given in Section \ref{Section_5}.

\section{Description of vehicle models for simulation or control purposes}
\label{Section_1}

The following section details some vehicle dynamics elements related to the longitudinal/lateral vehicle control problem. Three control approaches are developed: the first one is based on an algebraic nonlinear estimation and differential algebraic flatness of a 3DoF nonlinear two wheels vehicle model \eqref{Non_linear_bicycle_model}, the second one is based on simple PID controllers, in order to overcome the modeling problems, the last one is based on the so-called model-free control approach developped in \cite{ijc13}, since, under high dynamic loads, such models may become not sufficient to handle the critical maneuvers. For this purpose, the last model-free control law is developed to overcome the modeling problems. 

For vehicle simulation and control design problems, a large set of vehicle models exist. Usually, the linear bicycle model is widely used for control and estimation algorithms implementation. Furthermore, more complex models exist which are, however, not appropriated for such applications. For the flat nonlinear control issue, the following nonlinear coupled two wheels vehicle model is used (see e.g. \cite{Menhour13a}): 

\begin{equation}
\label{Non_linear_bicycle_model}
\left \{ \begin{array}{l}
m a_x=m(\dot{V}_x-\dot{\psi}V_y)= (F_{xf}+F_{xr}) \\
m a_y=m(\dot{V}_y+\dot{\psi}V_x)=(F_{yf}+F_{yr}) \\
I_z\ddot{\psi}=  L_fF_{yf} -L_rF_{yr}
\end{array}
\right.
\end{equation}

Notations are defined in Table \ref{notations_vehicle}. Notice that the the front and rear longitudinal forces in model \eqref{Non_linear_bicycle_model} are expressed using the following dynamical model of the tire forces:

\begin{equation}
\label{Front_rear_wheel}
\left \{ \begin{array}{lcl}
F_{xf} &= &(1/R)(- I_r\dot{\omega}_f + T_{m}-T_{bf})\\
F_{xr} &= &-(1/R)(T_{br} + I_r\dot{\omega}_r)
\end{array}
\right.
\end{equation}

The linear tire model of lateral forces in tha case of small slip angles is defined as follows:

\begin{equation}
\label{lin_lat_for}
\left \{
\begin{array}{l}
F_{yf}=C_f \left(\delta-\frac{V_y+\dot{\psi}L_f}{V_x}\right)\\[2mm]
F_{yr}=-C_r \left(\frac{V_y-\dot{\psi}L_r}{V_x}\right)
\end{array}
\right.
\end{equation}

\begin{figure*}[!ht]
\centering
\includegraphics[scale=0.6]{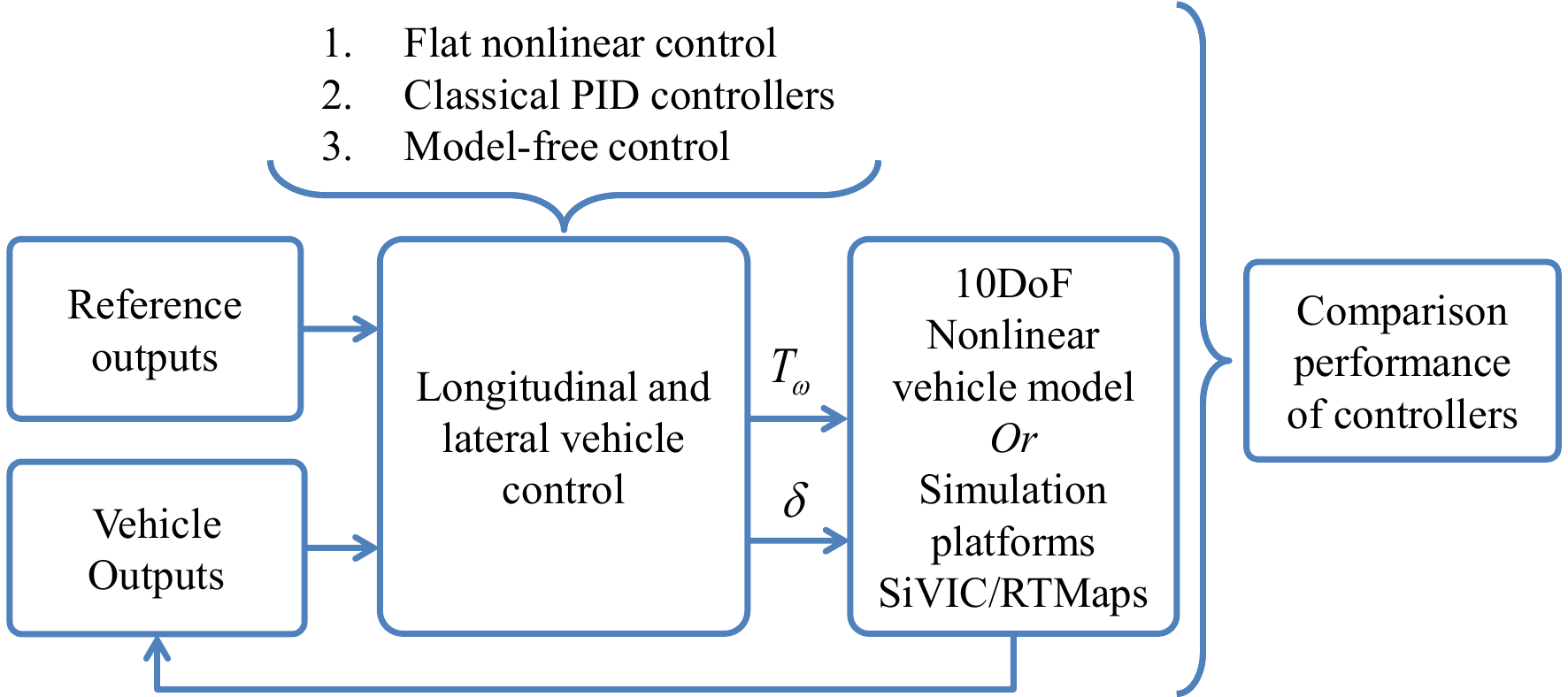}
\caption{Evolution of longitudinal and lateral vehicle control design and comparison}
\label{diag_valid_control_NB_V0}
\end{figure*}

For preliminary simulation results, a 10-DoF nonlinear vehicle model and real data recorded previously with a laboratory vehicle are used. The vehicle simulator considered here is a full nonlinear four wheels vehicle model \cite{Menhour13a} of a Peugeot 406 vehicle. Such a model is composed of longitudinal $V_x$, lateral $V_y$ and vertical $V_z$ translational motions, roll $\phi$, pitch $\theta$ and yaw $\psi$ rotational motions and dynamical models of the four wheels (see \cite{Menhour13a} for more details on the 10DoF nonlinear simulation vehicle model). All steps of design and simulation are summarized in Fig. \ref{diag_valid_control_NB_V0}.

In this document, the variables and notations in Table \ref{notations_vehicle}, with their meaning are used.

\begin{table}[th]
\caption{notations}
\label{notations_vehicle}
\centering
\begin{tabular}{cc}
\hline
Symbol & Variable name \\ \hline \hline
$V_x$ & longitudinal speed [$km.h$] \\ 
$V_y$ & lateral speed [$km.h$] \\ 
$\dot{\psi}$& yaw rate $[rad/s]$  \\ 
$\psi$ & yaw angle $[rad]$ \\ 
$y$ & lateral deviation $[m]$\\
$\dot{y}$ & derivative of lateral deviation $[m/s]$\\ 
$\beta$ & sideslip angle at the CoG $[rad]$ \\
$\delta$ & wheel steering angle $[rad]$ \\ 
$T_{\omega}$ & acceleration/braking torque $[Nm]$ \\
$F_{yf}$, $F_{yr}$ & front and rear lateral forces in the vehicle coordinate $[N]$ \\ 
$\alpha_f$, $\alpha_r$ & front and rear tire slip angles $[rad]$ \\ 
$\rho_x$ & aerodynamics drag coefficient\\
$C_f$, $C_r$ & front and rear cornering tire stiffnesses $[N/rad]$ \\ 
$L_f$, $L_r$ & distances from the CoG to the front and rear axles $[m]$ \\ 
$I_z$ & yaw moment of inertia $[kg.m^{-2}]$ \\ 
$I_x$ & moment of inertia about $x$ axis $[kg.m^{2}]$ \\ 
$m$ & vehicle mass $[kg m^2]$ \\ 
$g$ & acceleration due to gravity $[m/s^2]$ \\ 
$\mu$ & adhesion coefficient \\
\hline
\end{tabular}%
\end{table}

\section{Model-free control}
\label{Section_3}

\subsection{A background of the model-free control approach}

Model-free control was already applied and used quite successfully in a lot of various concrete examples (see the references in \cite{ijc13,ecc}). For obvious reasons let us insist here on its applications to \emph{intelligent transportation systems}: see \cite{Abouaissa12,Choi09,Villagra09,vil2},  and \cite{Menhour13b}. This last reference was briefly discussed in Section \ref{introd}.

\subsubsection{The ultra-local model}\label{A}

Replace the unknown SISO system by the \emph{ultra-local model}:
\begin{equation}
z^{(\nu)} = F + \alpha u
\label{ultralocal}
\end{equation}
where 
\begin{itemize}
\item $u$ is the control input,
\item $z$ is the controlled output,
\item $F$ is estimated via the measurements of the control input $u$ and the controlled output $z$. It does not distinguish between the unknown model of the system, the perturbations and uncertainties.
\item $\nu \geq 1$ is the derivation order,
\item $\alpha \in \mathbb{R}$ is chosen such that $\alpha u$ and $z^{(\nu)}$ are of the same order of magnitude, 
\item $\nu$ and $\alpha$ are chosen by the practitioner. 
\end{itemize}
\begin{remark}
In all the existing concrete examples 
$$\nu = \ 1 \ \text{or} \ 2$$ 
Until now from our knowledge, in the context of model-free control, the example of magnetic bearings \cite{Miras13} with their low friction, provides the only instance where the order $\nu = 2$ is necessary. 
\end{remark}

\subsubsection{Intelligent controllers}

Set $\nu = 2$ in Equation \eqref{ultralocal}:
\begin{equation}
\label{MFC_4_n_2}
\ddot{z} = F+ \alpha u 
\end{equation}
The corresponding \emph{intelligent Proportional-Integral-Derivative controller}, or \emph{iPID}, reads
\begin{equation}
\label{iPID_c}
u = - \frac{\left( F - \ddot{z}^d + K_P e + K_I\int e dt + K_D \dot{e}\right)}{\alpha}  
\end{equation}
where
\begin{itemize}	
\item $z^d$ is the desired reference trajectory,
\item $e = z-z^d$ is the tracking error and $z^d$ is a desired signal, 
\item $K_P$, $K_I$, $K_D \in \mathbb{R}$ are the usual gains.
\end{itemize}
Combining Equations \eqref{MFC_4_n_2} and \eqref{iPID_c} yields
$$
\ddot{e} + K_P e + K_I \int e dt + K_D \dot{e}= 0
$$
where $F$ does not appear anymore. Gain tuning becomes therefore quite straightforward. This is a major benefit when compared to ``classic'' PIDs. 
If $K_I = 0$ we obtain the \emph{intelligent Proportional-Derivative controller}, or \emph{iPD},
\begin{equation}
\label{iPD_c}
u = - \frac{\left( F - \ddot{z}^d + K_P e + K_D \dot{e}\right)}{\alpha}  
\end{equation}
Set $\nu = 1$ in Equation \eqref{ultralocal}:
\begin{equation}
\label{MFC_4_n_1}
\dot{z} = F+ \alpha u 
\end{equation}
The corresponding \emph{intelligent Proportional-Integral controller}, or \emph{iPI}, reads: 
\begin{equation}
\label{iPI_c}
u = - \frac{\left(F - \dot{z}^d + K_P e + K_I\int e dt \right)}{\alpha}
\end{equation}	
If $K_I = 0$ in Equation \eqref{iPI_c}, we obtain the \emph{intelligent proportional controller}, or \emph{iP}, which turns out until now to be the most useful intelligent controller:
\begin{equation}
\label{ip}
{u = - \frac{F - \dot{z}^d + K_P e}{\alpha}}
\end{equation} 

\subsubsection{Algebraic estimation of $F$}
\label{F}

$F$ in Equation \eqref{ultralocal} is assumed to be ``well'' approximated by a piecewise constant function $F_{\text{est}} $. According to the algebraic parameter identification developed in \cite{sira1,sira2}, where the probabilistic properties of the corrupting noises may be ignored, if $\nu = 1$, Equation \eqref{MFC_4_n_1} rewrites in the operational domain (see, \textit{e.g.}, \cite{Yosida84}) 
$$
s Z = \frac{\Phi}{s}+\alpha U +z(0)
$$
where $\Phi$ is a constant. We get rid of the initial condition $z(0)$ by multiplying the both sides on the left by $\frac{d}{ds}$:
$$
Z + s\frac{dZ}{ds}=-\frac{\Phi}{s^2}+\alpha \frac{dU}{ds}
$$
Noise attenuation is achieved by multiplying both sides on the left by $s^{-2}$. It yields in the time domain the realtime estimation

\begin{equation}
\label{integral}
 F_{\text{est}}(t)  =
-\frac{6}{\tau^3}\int_{t-\tau}^t \left\lbrack 
\begin{array}{c}
(\tau -2\sigma)z(\sigma)\\
+\alpha\sigma(\tau -\sigma)u(\sigma) 
\end{array}
\right\rbrack d\sigma 
\end{equation}

Notice that the extension to the case $\nu = 2$ is straightforward. For this case, the estimation of $F$ is performed by the following estimator:

\begin{equation}
\label{integral_2}
F_{\text{est}}(t) = 
\left[
\begin{array}{c}
-\frac{60}{\tau^5}\int_{t-\tau}^t(\tau^2 +6\sigma^2 - 6\tau \sigma)z(\sigma)d\sigma \\[2mm]
-\frac{30\alpha}{\tau^5} \int_{t-\tau}^t(\tau - \sigma)^2\sigma^2 u(\sigma)d\sigma
\end{array}
\right]
\end{equation}
where $\tau > 0$ might be quite small. This integral may, of course, be replaced in practice by a classic digital filter (for more details see e.g. \cite{ijc13}). See \cite{nice} for a cheap and small hardware implementation of our controller. It should be emphasized that the above-mentioned estimation methods are not of asymptotic type. 

\subsection{Application to a vehicle control}

As previously mentioned, to take advantage of model-free approach, an appropriate choice of the outputs and the corresponding inputs control is required. Since to avoid any vehicle modeling problem and ensure a desired tracking longitudinal and lateral motions, the following input and output variables are selected: 

\begin{enumerate}
\item the acceleration/braking torque $u_1 = T_\omega$ and the longitudinal speed $z_1$,
\item the steering wheel angle  $u_2 = \delta$ and the lateral deviation $z_2$.
\end{enumerate}

It is obvious that the second output, which is the lateral deviation output, gives a kinematic relationship between the vehicle motions like longitudinal, lateral and yaw motions. This allow us to include some coupling effects between these motions. According to the background on the model-free setting, the above inputs/outputs, and the Newton's second law, the following two local models are deduced:

 \begin{eqnarray}
\label{MFC_LLVC2_1}
\text{longitudinal local model:} & \dot{z}_1 = F_1+ \alpha_1 u_1\\
\label{MFC_LLVC2_2}
\text{lateral local model:} & \ddot{z}_2 = F_2+ \alpha_2 u_2 
\end{eqnarray}
Note the following properties:
\begin{itemize}
\item Equations \eqref{MFC_LLVC2_1}-\eqref{MFC_LLVC2_2} seem decoupled, but the coupling effects are included in the terms $F_1$ and $F_2$.
\item Equation \eqref{MFC_LLVC2_2} is an order $2$ formula with respect to the derivative of $z_2$.
\end{itemize}
For Equation \eqref{MFC_LLVC2_1} (resp. \eqref{MFC_LLVC2_2}), the loop is closed by an iP \eqref{ip} (resp. iPD \eqref{iPD_c}) as follows:
 \begin{eqnarray}
\label{MFC_Control_1}
u_1 & = & -\frac{1}{\alpha_1} \left( F_1 - {\dot{z}}_1^{d} + K_P^{z_1} e_{z_1}  \right)\\
\label{MFC_Control_2}
u_2 & =  & -\frac{1}{\alpha_2}\left( F_2 - {\ddot{z}}_2^{d} + K_P^{z_2}e_{z_2} + K_D^{z_2}\dot{e}_{z_2} \right)
\end{eqnarray}

Moreover, the estimators \eqref{integral} and \eqref{integral_2} are powerful tools to estimate respectively $F_1$ and $F_2$ as follows:
\begin{itemize}
	\item For $\nu = 1$
\begin{equation}
\label{integral_app}
 F_{1_\text{est}}(t)  =
 -\frac{6}{\tau^3}\int_{t-\tau}^t \left\lbrack
 \begin{array}{c}
  (\tau -2\sigma)z_1(\sigma)\\
  +\alpha\sigma(\tau -\sigma)u_1(\sigma) 
   \end{array}
  \right\rbrack d\sigma 
\end{equation}

	\item For $\nu = 2$
\begin{equation}
\label{integral_2_app}
F_{2_\text{est}}(t) = 
 \left[
\begin{array}{c}
-\frac{60}{\tau^5}\int_{t-\tau}^t(\tau^2 +6\sigma^2 - 6\tau \sigma)z_2(\sigma)d\sigma\\
-\frac{30\alpha}{\tau^5} \int_{t-\tau}^t(\tau - \sigma)^2\sigma^2 u_2(\sigma)d\sigma
 \end{array}
\right]
\end{equation}
\end{itemize}

Table \ref{Gains_MFC} gives the gains of the model-free control used in Section \ref{Section_4}. 

\begin{table}[!ht]
\caption{Model-Free Control parameters}
\label{Gains_MFC}
\centering
\begin{tabular}{|c|c|c|c|c|c|}
\cline{2-6}
 	 \multicolumn{1}{c|}{}					& $\alpha$ 		& $K_P$ 	& $K_D$ 					& $\tau\, [ms]$ 	& f\,[Hz]	\\ \hline 
   Longitudinal control $u_1$ 		& 1.5 				&  2 			& 0						 	  &  0.25  					& 200 		\\ \hline 
   Lateral control $u_2$ 					& 1.95 				&  1.9		&  0.5 						&  0.25  					& 200			\\  \hline
  \end{tabular}
\end{table}

\subsection{Lateral deviation calculation}
\label{Section_42}

The parameters that easily characterize a desired reference trajectory are the road curvature $\rho_d$, the yaw angle $\psi_d$, the path length coordinate $s_d$, the coordinates $x_d$ and $y_d$. The road curvature is computed by longitudinal speed and lateral acceleration using the following formula:

\begin{equation}
\label{corbure}
\rho_d(s_d)=\frac{1}{R(s_d)}=\frac{a_y(s_d)}{V_x^2(s_d)}
\end{equation}

This method computes the yaw angle from the curvature \eqref{corbure}, then the coordinates $\psi_d$, $x_d$ and $y_d$ are deduced as follows by projection of the path length coordinate $s_d$ in the vehicle frame ($s_0$ being the initial condition):
%
\begin{equation}
\label{Model_trajectory}
\left\{
\begin{array}{l}
\psi_{d}(s_{d}) =  \int_{s_{0}}^{s}{\rho_{d}(s_{d})ds} \\
x_{d}(s_{d})=  \int_{s_{0}}^{s}{\cos(\psi_d(s_{d}))ds}\\
y_{d}(s_{d})= \int_{s_{0}}^{s}{\sin(\psi_d(s_{d}))ds}
\end{array}\right.
\end{equation}

\begin{figure}[!ht]
\centering
\includegraphics[scale=0.55]{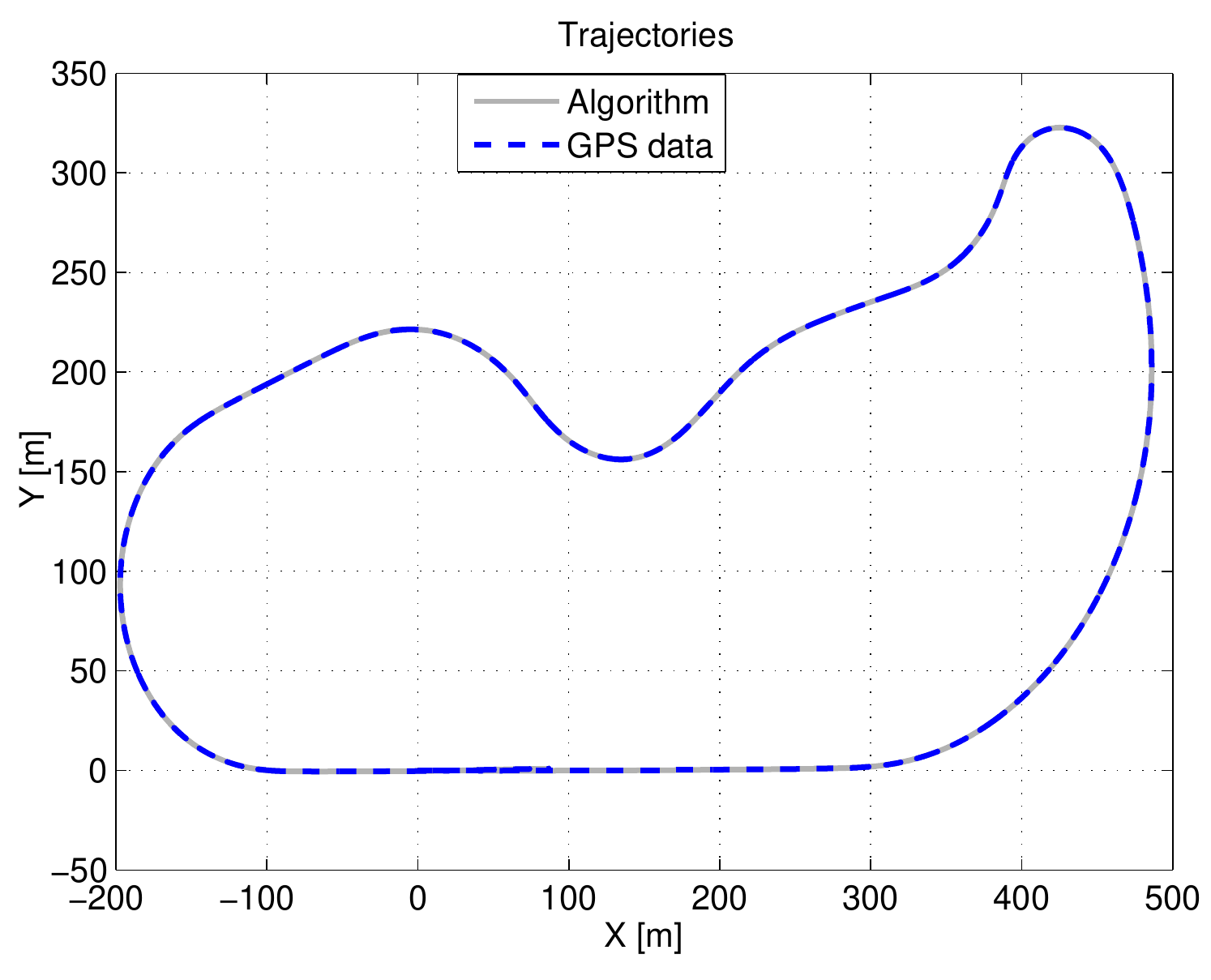}
\caption{GPS trajectory and trajectory of the algorithm \eqref{Model_trajectory}}
\label{Trajectories_Model_GPS}
\end{figure}

Fig. \ref{Trajectories_Model_GPS} shows an experimental validation of algorithm \eqref{Model_trajectory}. The actual GPS data (grey curve) are quite close to the computed trajectory (dash blue curve). Moreover, this algorithm only depends on measured data and it is independent of any vehicle parameters.

Unfortunately, without accurate ground truth requiring very expensive embedded means and sensors, it is very difficult to quantify precisely the impact and the quality of the obtained result. In the real experiments, we have observed that the results were interesting and relevant. In order to obtain an objective view of the level of quality of the proposed model-free approach comparatively to classical one, the use of simulation platforms is performed.

\section{Simulation results}
\label{Section_4}

As we mentioned previously in this paper, actual condition tests often are expensive and require specially equipped vehicles in order to study the interactions between embedded applications and events encountered in the road environment. Moreover in actual conditions, tens of thousands of kilometers are required in order to obtain good enough estimation of an application quality and robustness. Furthermore, with the development of autonomous, cooperative and connecting systems, applications become increasingly complex to implement and validate. Then the Simulation allows to approach the reality of a situation and allows the control of specific events and their reproducibility. The simulation allows too to play with weather factors and to study the influence and the impact of physical parameters on the robustness of the functions used in embedded applications, especially connected and active ones (cooperative and automated driving assistance systems). Difficult or dangerous tests are easily accessible to simulation as well as the analysis of the functions performance in a singular event (emergency braking or road departure warning due to obstacle detection or road friction problem), whereas in real condition, waiting hours could be required before the generation of such an event. Waiting for this real event would be inefficient and unreasonable.

\begin{figure*}[!ht]
\centering
\includegraphics[scale=0.8]{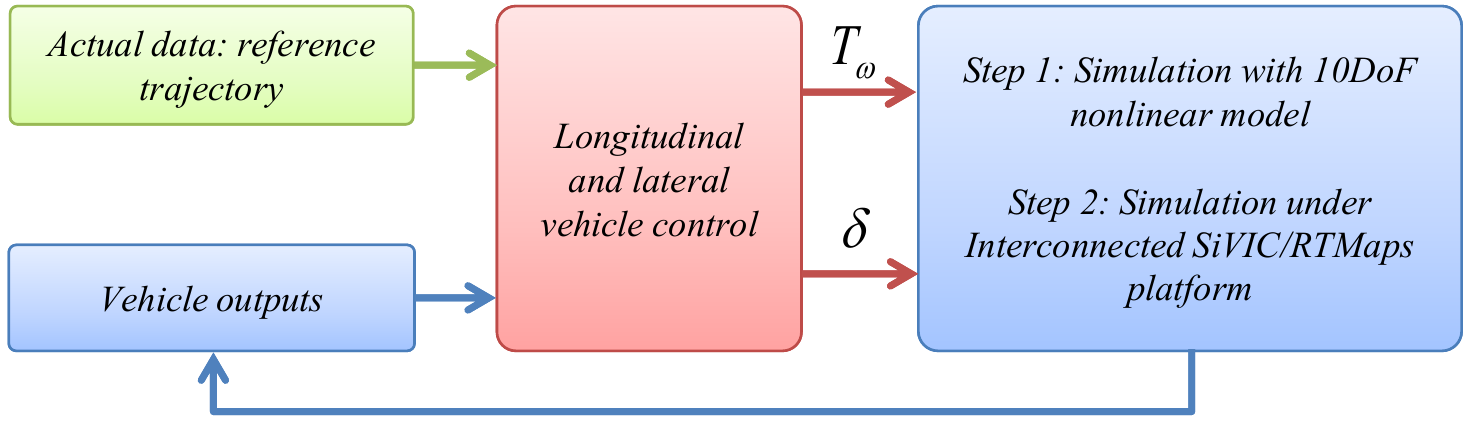}
\caption{Validation diagram block of closed-loop system}
\label{diag_valid_control_NB_V1}
\end{figure*}

In this study, the simulation stage is carried out according to the diagram block of Fig. \ref{diag_valid_control_NB_V1} and uses two levels of simulation and three different platforms: firstly a Matlab platform is used with a full nonlinear model of the instrumented Peugeot 406 car as in \cite{Menhour13b}, and secondly thanks to the pro-SiVIC\footnote{pro-SiVIC is a professional software of CIVITEC ({\tt http://www.civitec.com}).} platform \cite{Gruyer10b, Gruyer10a, Vanholme10, Gruyer09} interconnected with RTMaps platform\footnote{RT-Maps is developed by Intempora ({\tt http://www.intempora.com}).}, we will simulate complex vehicle dynamical modeling \cite{IJACSA13}, environment, infrastructure, and realistic embedded sensors. Moreover pro-SiVIC will provide mechanisms allowing to generate very accurate ground truth for the evaluation and validation stages. RTMaps is a platform which allows to record, replay, manage and process multiple data flows in real-time.

\subsection{Simulation under Matlab}

\begin{figure*}[!ht]
\centering
\includegraphics[scale=0.53]{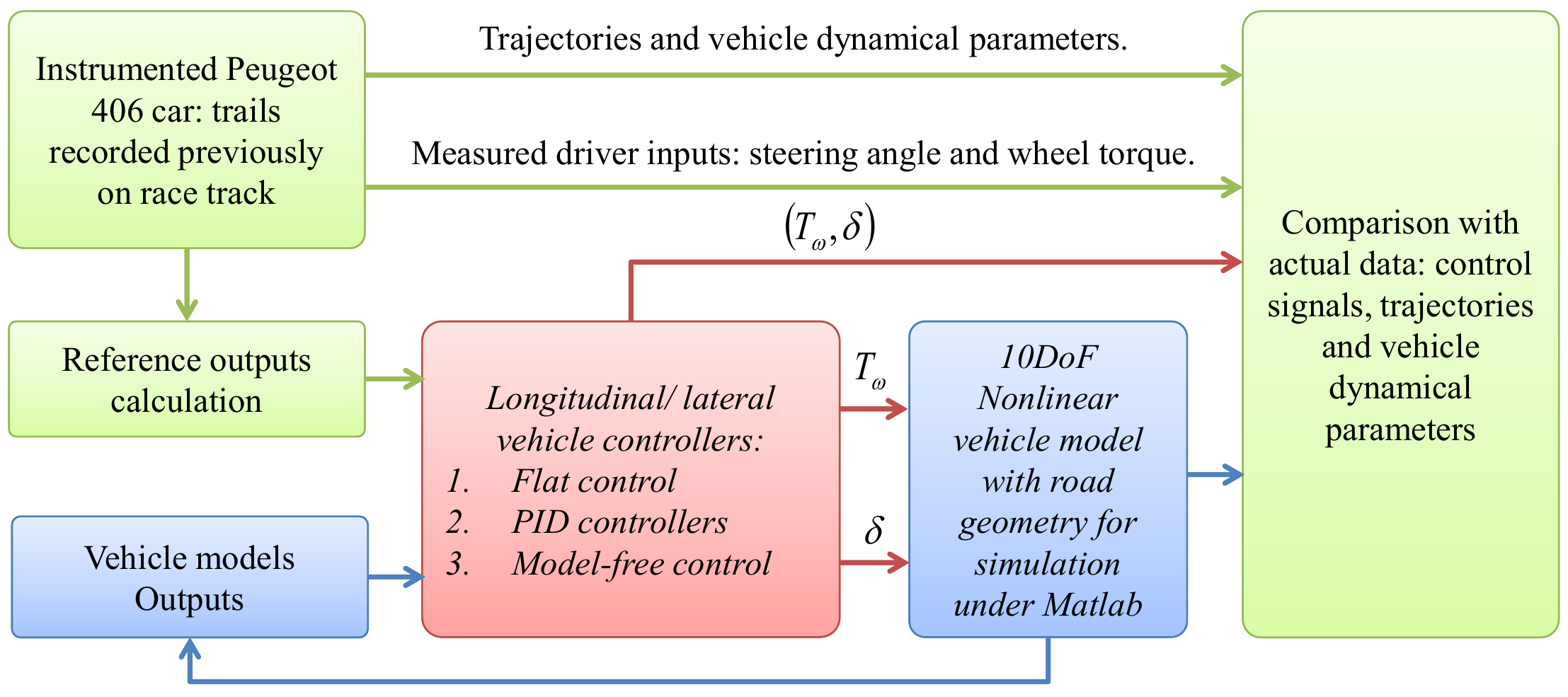}
\caption{Validation diagram block of vehicle controllers, reference trajectory reconstruction and vehicle models}
\label{diag_valid_control_NB_V2}
\end{figure*}

The first simulation is conducted according to the diagram block of Fig. \ref{diag_valid_control_NB_V2} using a 10DoF nonlinear vehicle model\footnote{See \cite{Menhour13a} and Section \ref{Section_1} for more details on 10DoF nonlinear vehicle model} and real data recorded previously with a prototype vehicle from CAOR laboratory (Mines Paristech). Several experiments have been realized with an instrumented car to compute the coordinates of the race track depicted in Figure \ref{Trajectories_Model_GPS}. For each trial, several dynamical variables have been recorded; among them: lateral and longitudinal accelerations, longitudinal and lateral speeds, yaw and roll rates, wheel rotation speeds, moments on the four wheels, longitudinal, lateral and vertical forces on the four wheels, steering angle, etc. Moreover, the road geometry (road bank and slope angles) is considered in the closed-loop system. In these simulations, the efficiency and relevance of the model-free control as well as its performances are compared to those of nonlinear flat control and PID controllers. The Table \ref{Inputs_outputs} summarizes the control inputs and the controlled outputs used for each control law. Two simulation scenarios are conducted for two values of the road friction coefficient: the first one for dry asphalt $\mu = 1$ and the second one for wet asphalt $\mu = 0.7$. These scenarios mean that the adhesion capability of the ground is reduced, thus, the vehicle becomes unstable, i.e that the vehicle maneuverability and controllability become more difficult.

\begin{table*}[!ht]
\caption{control inputs and controlled outputs of each control law}
\label{Inputs_outputs}
\centering
\begin{tabular}{|c|c|c|c|c|c|}
\cline{3-5}
 \multicolumn{2}{c|}{}               & \multicolumn{1}{c|}{PID control}    &  \multicolumn{1}{c|}{Flat control}   & \multicolumn{1}{c|}{MFC}\\
   \hline 
   \multirow{2}{*}{{Inputs}}  & Driving/braking torque  $T_\omega$  &  $\times$ &  $\times$ &   $\times$ \\ 
   \cline{2-5} 
   \multicolumn{1}{|c|}{}            & Steering angle $\delta$             & $\times $ &  $\times$ &   $\times$ \\
   \hline
   \hline
   \multirow{3}{*}{{Outputs}} & Longitudinal speed $V_x$            &  $\times$ &  $\times$ &  $\times$ \\ 
   \cline{2-5}                       & Second flat output $z_2$            &  	NA					&  $\times$ &          NA    \\ 
   \cline{2-5}                       & Lateral deviation $y$               &  $\times$ &      NA        &  $\times$ \\ 
   \hline
  \end{tabular}
\end{table*}

\begin{figure}[!ht]
\centering
\includegraphics[scale=0.58]{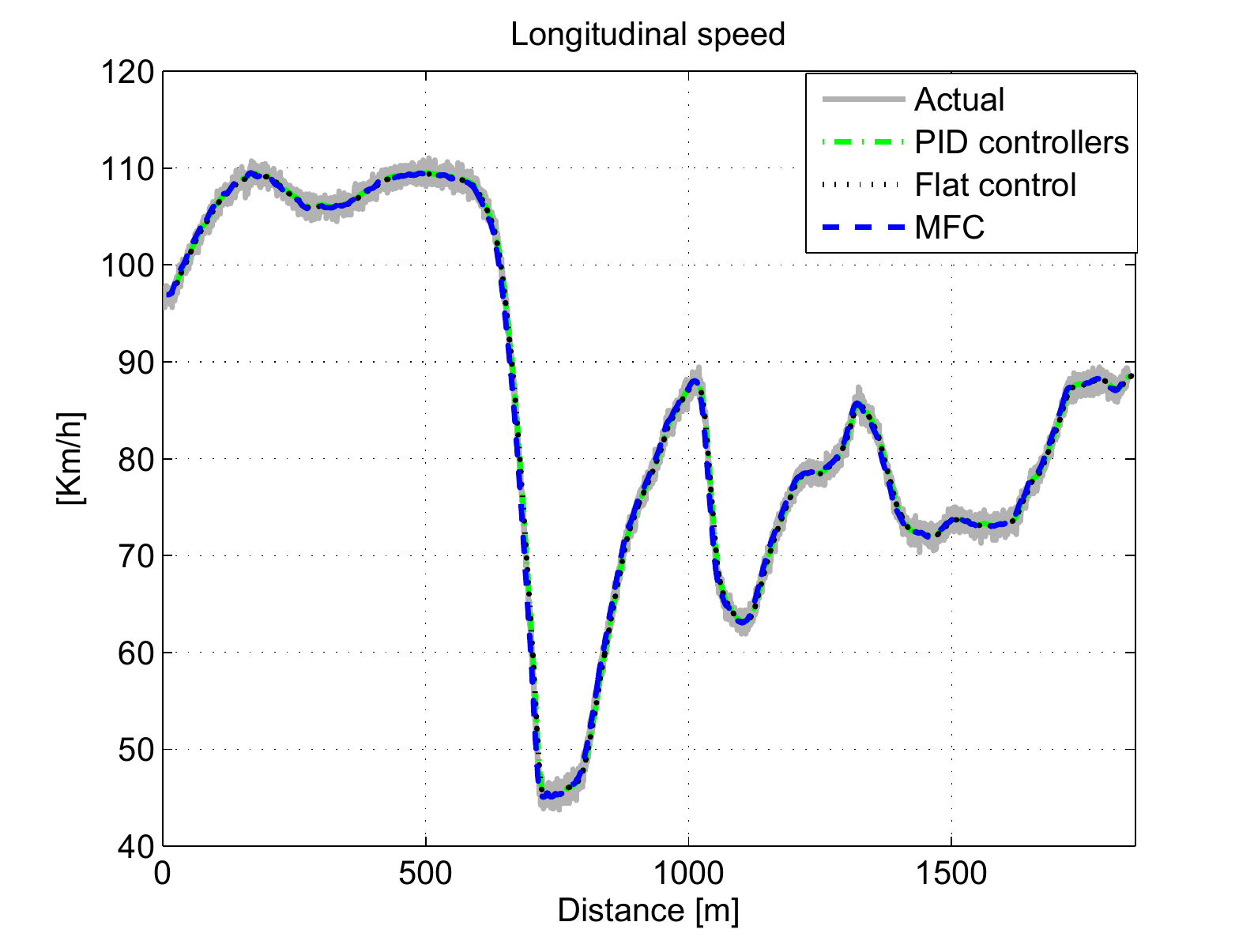}
\caption{The actual longitudinal speed versus the closed-loop simulated longitudinal velocities}
\label{Vx_comp}
\end{figure}

\begin{figure}[!ht]
\centering
\includegraphics[scale=0.58]{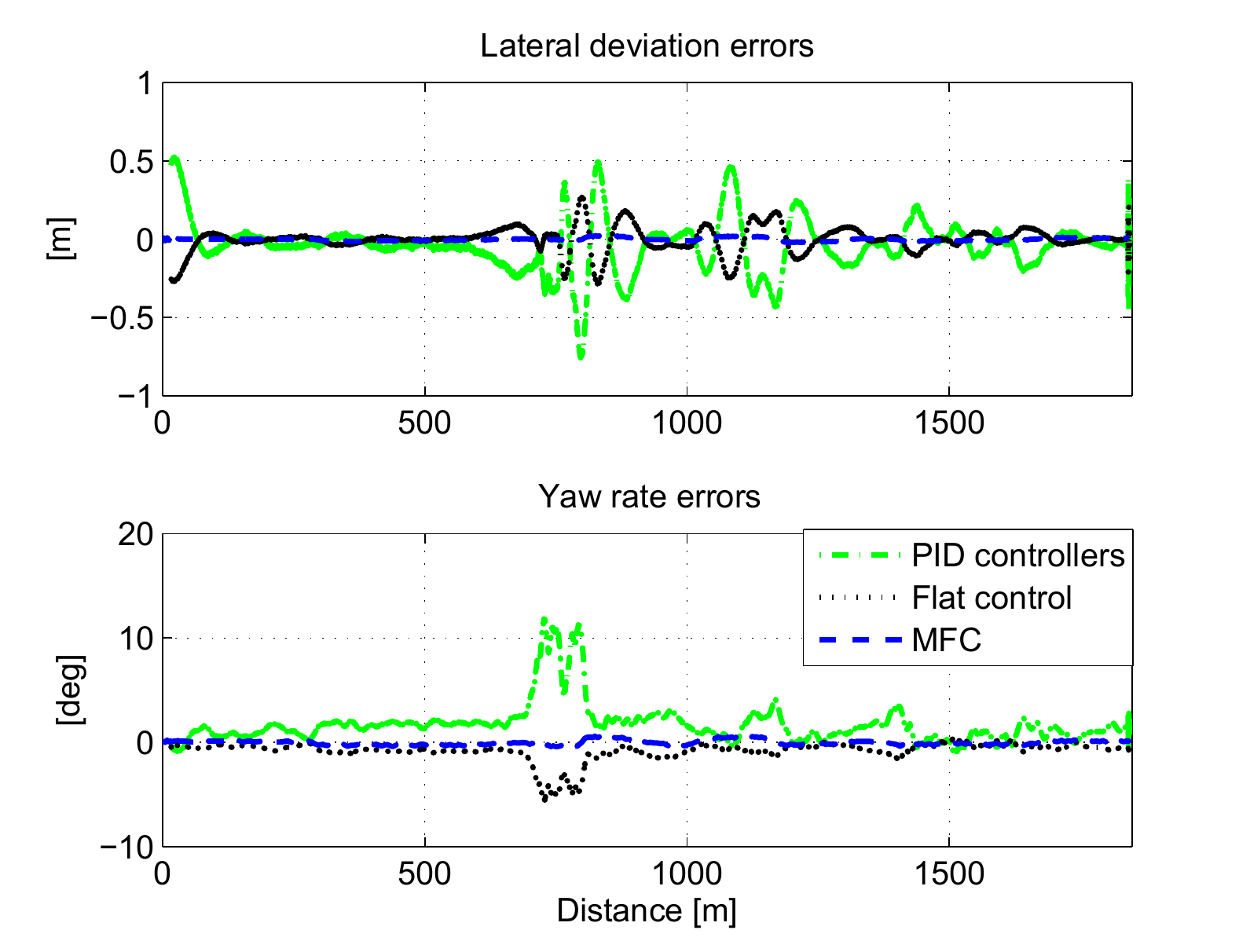}
\caption{Tracking trajectory errors on lateral deviation and yaw angle}
\label{Ey_Epsi}
\end{figure}

\begin{figure}[!ht]
\centering
\includegraphics[scale=0.58]{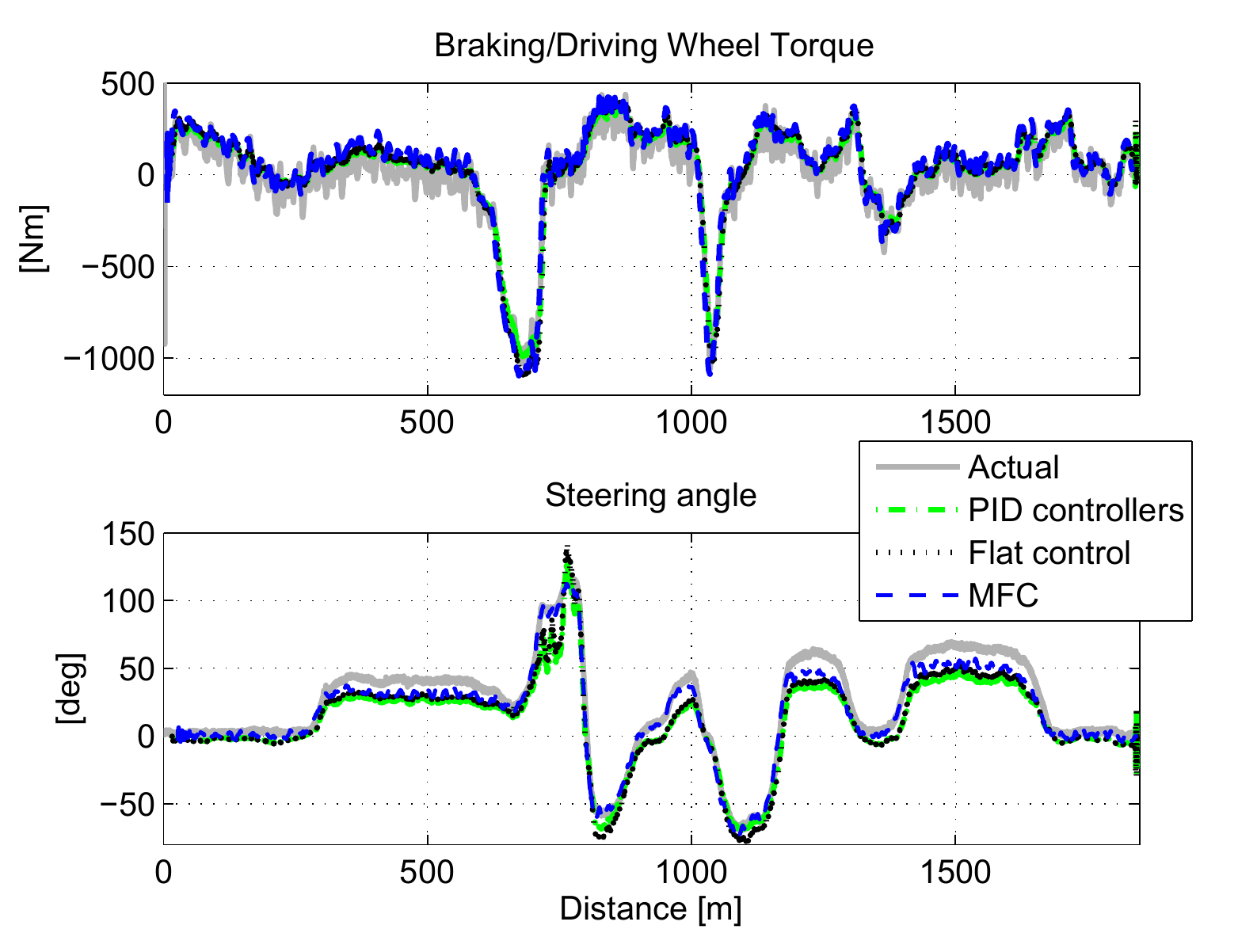}
\caption{Wheel torques and steering angles control signals: actual and simulated}
\label{Twheel_Delta}
\end{figure}

For a dry road ($\mu = 1$), results are shown in Figs.  \ref{Vx_comp}, \ref{Ey_Epsi} and \ref{Twheel_Delta}. These simulations demonstrate that the model-free control (MFC) gives quite satisfying results, better than those obtained with PID control and nonlinear flat control. It should be noticed that the test track which has been considered implies strong lateral and longitudinal dynamical requests. This track involves different types of curvatures linked to straight parts, and all these configurations represent a large set of driving situations. Fig. \ref{Vx_comp} shows that all controllers produce accurate behavior for autonomous driving applications. It is obvious that the results displayed on Fig. \ref{Ey_Epsi}, show that the tracking errors on the lateral deviation and on the yaw angle outputs produced by the model-free control (MFC) are better compared to those produced by the other controllers. These errors are less than $10\, cm$ and $0.5\, deg$ in the case of model-free control (MFC). Finally, Fig. \ref{Twheel_Delta} shows that the control signals computed by all control strategies are quite close to the actual ones provided by the driver along the track race. 

Moreover, the effectiveness of the model-free control (MFC) results is also evaluated through the following normalized error:

\begin{equation}
\label{norma_errors}
e_z(i)={100 \left|z_{s}(i)-z_{act}(i) \right|}/{ \max \left| z_{act} \right|}
\end{equation}

\begin{table*}[!ht]
\caption{Maximum normalized errors on some variables of the closed-loop system for dry and wet road}
\label{errors_dyn_para}
\centering
\begin{tabular}{|c|c|c|c|c|c|}
\cline{3-5}
 \multicolumn{2}{c|}{} &\multicolumn{1}{c|}{PID control} &\multicolumn{1}{c|}{Flat control} & \multicolumn{1}{c|}{MFC}\\
      \hline
   \multirow{3}{*}{\begin{sideways}{$\mu = 1$}\end{sideways}} & Longitudinal speed $e_{V_x}$ (\%) &  0.93 &  0.45 & 0.186 \\ 
   \cline{2-5}
& Yaw angle $e_{\psi}$ (\%)    & 1.76 &  1.21 & 0.45 \\  
  \cline{2-5}
& Lateral deviation $e_{y}$ (\%)  & 2.8 &  1.4 &  0.35 \\  
   \hline
   \hline
   %
      \multirow{3}{*}{\begin{sideways}{$\mu = 0.7$}\end{sideways}}
       & Longitudinal speed $e_{V_x}$ (\%)  &  5.54 &  4.23  & 2.31 \\ 
   \cline{2-5}
&Yaw angle $e_{\psi}$ (\%)  & 13.54 & 7.67  & 2.7\\ 
\cline{2-5}
& Lateral deviation $e_{y}$ (\%) &  16.64 & 9.37 &  3.49 \\ 
   \hline
  \end{tabular}
\end{table*}

Table \ref{errors_dyn_para} summarizes a comparison between the closed-loop simulated results $z_{s}$ and the actual data $z_{act}$ using the normalized error \eqref{norma_errors}. The normalized norm is computed for dry and wet road scenarios. In the two cases, the model-free control (MFC) provides better results than PID control and nonlinear flat control. Using the model-free control (MFC), the maximum normalized error values are less than $0.5 \,\%$ for dry road test and less than $3.5 \,\%$ for wet road test. However, all normalized errors obtained with other controllers are deteriorated mainly with wet road test. These results confirm that the model-free control (MFC) with natural outputs produces a satisfactory behavior. 

\subsection{Simulation under pro-SiVIC interconnected with RTMaps}
\begin{figure*}[!ht]
\centering
\includegraphics[scale=0.38]{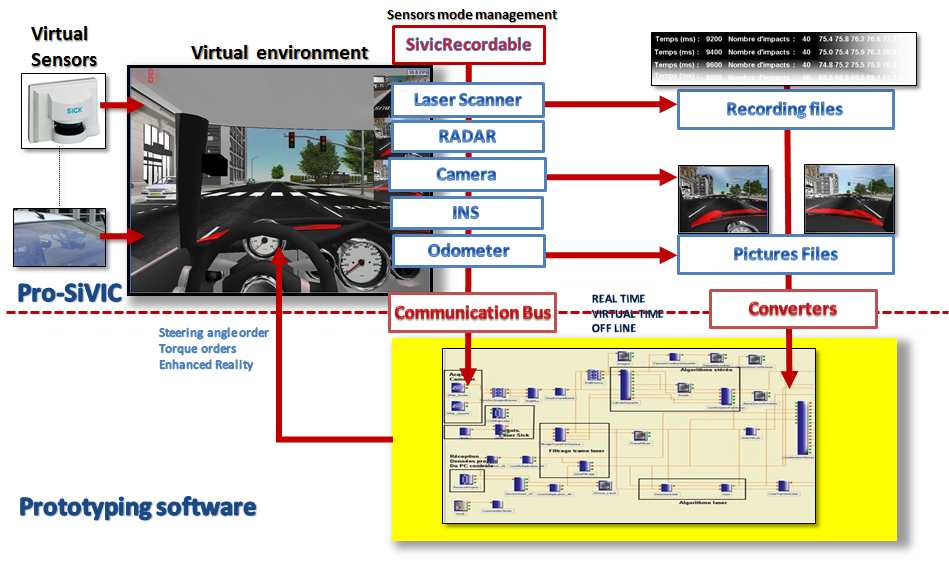}
\caption{pro-SiVIC/RTMaps interconnected platforms for real-time prototyping, test and validation}
\label{SiVIC_RTMaps_architecture}
\end{figure*}

After the Matlab stage for the model-free approach validation, a second simulation level is addressed. This second stage proposes to implement the model-free control algorithms in a more realistic way. In this case, very realistic and complex vehicle modeling is used. This vehicle level of complexity is presented in \cite{IJACSA13}. Attached to this vehicle model, a set of virtual sensors are simulated. All this functionalities are available in pro-SiVIC platform. The data generated by pro-SiVIC and its sensors are sent to RTMaps which is the real software environment used in a real prototype vehicle in order to implement ADAS (Advanced Driving Assistance Systems) applications. Using an efficient way to interconnect these 2 platforms, we obtain an efficient Software In the Loop (SIL) solution and a full real-time simulation architecture presented in Fig. \ref{SiVIC_RTMaps_architecture}. 

\subsubsection{The pro-SiVIC platform}
From the past decade, a great set of researches aimed to improve the road safety through the implementation of driving assistance systems. These researches generally take into account a local perception (from the ego-vehicle point of view) and the ego-vehicle maneuvers (e.g. braking and acceleration). Nevertheless, in many actual road configurations, a local perception is no longer sufficient. Additional information is needed to minimize risk and maximize the driving security level. This additional information requires additional resources which are both time-consuming and expensive. It therefore becomes essential to have a simulation environment allowing to prototype and to evaluate extended, enriched and cooperative driving assistance systems in the early design stages. An efficient and functional simulation platform has to integrate different important capabilities: Road environments, virtual embedded sensors (proprioceptive, exteroceptive), infrastructure sensors and cooperative devices (transponders, communication means, ...), according to the physical laws. In the same way, a physics-based model for vehicle dynamics (potentially with steering wheel column and powertrain models) coupled with actuators (steering wheel angle, torques on each wheel) is necessary. pro-SiVIC takes into account these remarks and these requirements, and is therefore a very efficient and dedicated platform to develop and prototype a high level autonomous driving system with cooperative and extended environment perception. 

\begin{figure*}[!ht]
\centering
\includegraphics[scale=0.7]{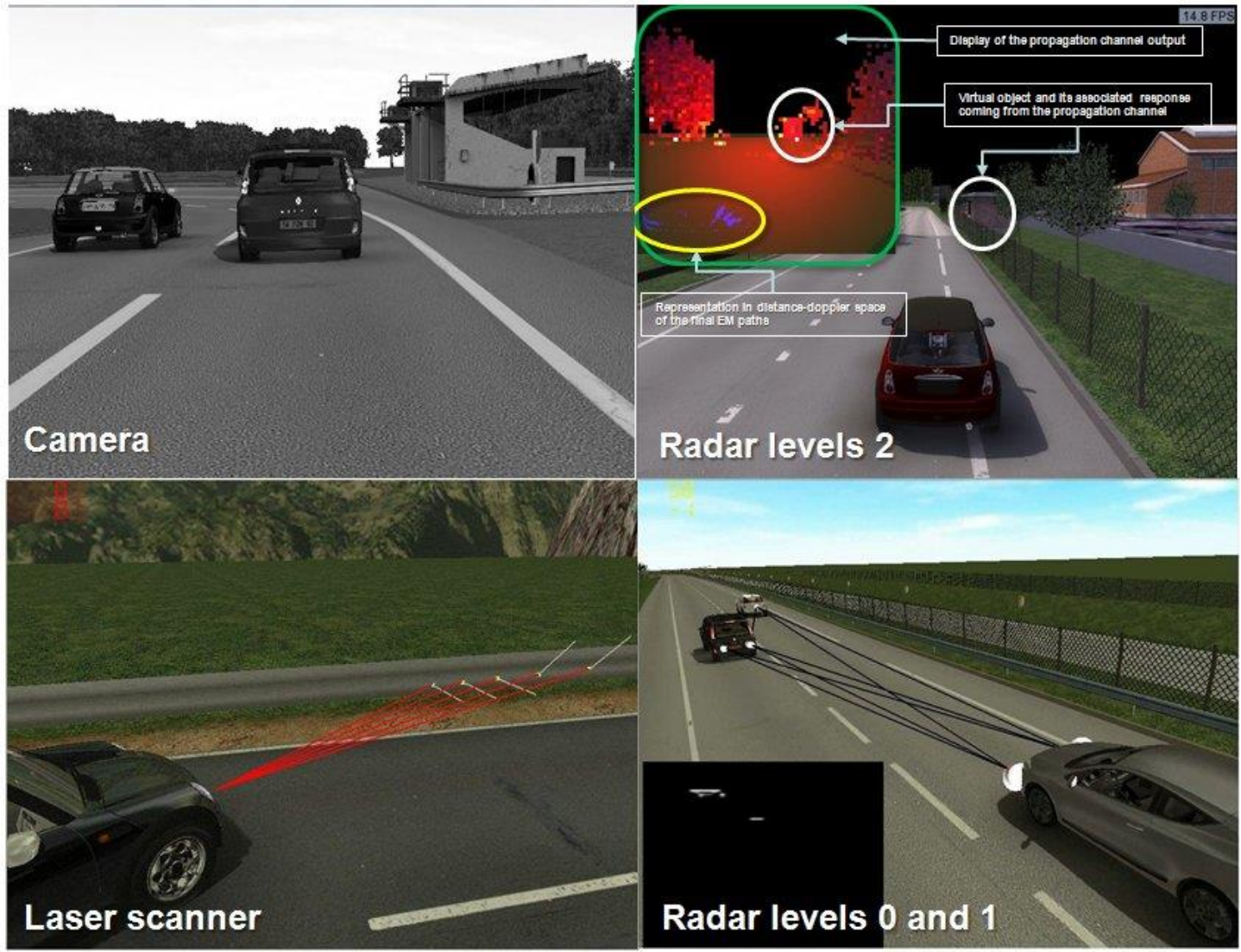}
\caption{Some exteroceptive types of sensors in pro-SiVIC}
\label{SiVIC_Sensors}
\end{figure*}

This platform is currently shared in two version: pro-SiVIC and pro-SiVIC research. The first version is the industrial one. The second version, which we have used, is the research version developed in IFSTTAR institute. The research version, in its current state, includes a set of different exteroceptive and proprioceptive sensors, thus communication means. The exteroceptive sensors set involved several types of cameras (classical, omnidirectional, and fisheye), several models of laser scanner, and some levels of automotive RADAR modeling. The proprioceptive sensors model odometers (curvilinear distance, speed, impulses) and Inertial Navigation systems (accelerometers, gyrometers). Then communication means for cooperative systems include both 802.11p communication media and beacon (transponder). For all these sensors and medias, it is possible, in real-time and during the simulation stages, to tune and to change the sensor's sampling time, operating modes, and intrinsic/ extrinsic parameters. Some examples of the rendering of these sensors are shown in Fig. \ref{SiVIC_Sensors}.

\subsubsection{RTMaps platform}
RTMaps is a software platform originally developed at Mines ParisTech by B. Steux \cite{Steux01}.\footnote{It is now marketed by the company Intempora.} for the needs of the European Carsens project. The main objective of this platform was to provide an efficient way to collect, to replay and to process, in real real-time, a great set of heterogeneous and asynchronous sensors. So this platform can provide a relevant solution for the embedded and multi-sensors applications prototyping, test and evaluation. In the processing stages, RTMaps allows to manage and to process in same time raw data flows coming from images, laser scanner, GPS, odometric, and INS sensors. The algorithms, which can be applied to the sensor data, are provided in several libraries (RTMaps packages) dedicated to specific processing (image processing and multi-sensors fusion). In its current user configuration an efficient new package and new module development procedure is available. This capability has been used in order to implement our different perception and control modules. This type of architecture gives a powerful tool in order to prototype embedded ADAS (Advanced Driving Assistance Systems) with either informative outputs or orders to control vehicle dynamics. At each stage, the sensor data and module outputs are time-stamped for an accurate and a reliable time management.

\subsubsection{pro-SiVIC/RTMaps: an interconnected platform for efficient Advanced Driving Assistance Systems prototyping}
\begin{figure*}[!ht]
\centering
\includegraphics[scale=0.65]{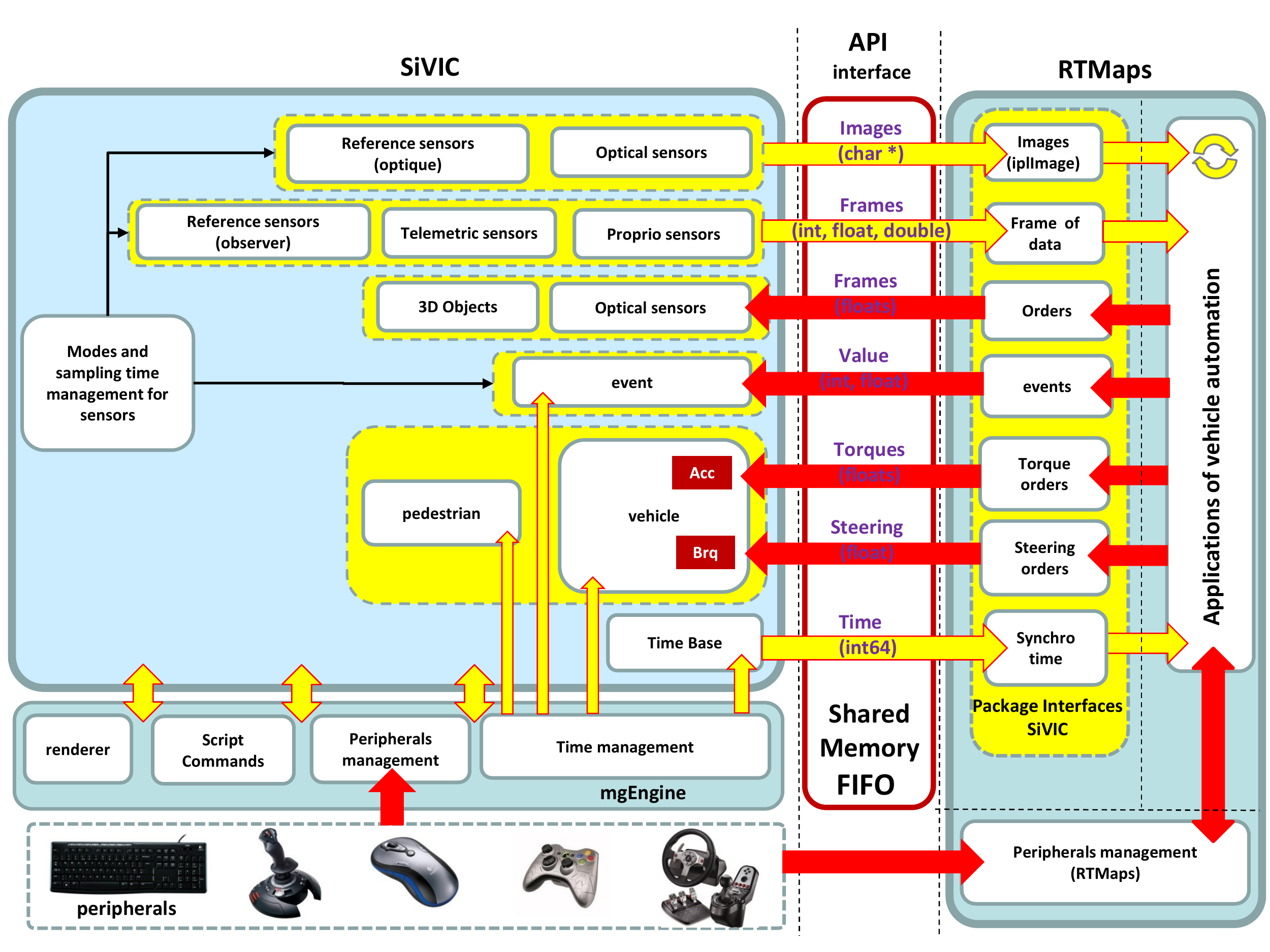}
\caption{SiVIC/Types of data managed between pro-SiVIC and RTMaps}
\label{SiVIC_Interconnections}
\end{figure*}

The interconnection between pro-SiVIC and RTMaps platforms brings to RTMaps the ability to replace real raw data coming from embedded sensors by simulated ones. This platforms coupling provides an efficient and a solid framework for the prototyping and the evaluation of control/command and perception algorithms dedicated to autonomous driving and SIL applications. In this multiple platform, the raw data coming from virtual sensors, vehicles, and ground truth are sent to RTMaps with, for each one, a time stamp. These data are used as inputs in the applications developed in a specific package involving a perception/control/command module. Then the outputs of the control/command module provide the orders which are sent back to the interface module which allows, in pro-SiVIC, to control the longitudinal and lateral maneuvers of the virtual ego-vehicle. In our case, we do not use the powertrain and steering wheel models. This implies that the orders act directly on the dynamics vehicle model (front wheel angle and wheel torques). This chain of design is very efficient because the algorithms developed in RTMaps can then be directly transferred either as a micro-software on real hardware devices, or in an embedded version of RTMaps in an actual prototype vehicle. In this way and with the high level of vehicle and sensors modeling, the design process can be considered very close to reality (real vehicles, real sensors). The different modes, flow of data, types of data, type of peripherals handled by this interconnection mechanism are shown in Fig. \ref{SiVIC_Interconnections}.

Several mechanisms have been implemented and tested. The best solution is clearly the optimized FIFO (First In, First Out) method which allows the transfer of a great number of data in a short time. It is a very critical functionality in order to guarantee a real-time link between pro-SiVIC and the perception/data processing/control algorithms. In order to correctly manage time, a synchronization module is available. This synchronization allows providing a time reference from pro-SiVIC to RTMaps. Then RTMaps is fully synchronized with pro-SiVIC components (vehicle, pedestrian and sensors). The pro-SiVIC/RTMaps simulation platform also enables to build reference scenarios and allows evaluating and testing of control/command and perception algorithms. In fact, the pro-SiVIC/RTMaps platform constitutes a full simulation environment because it provides the same types of interactivity found on actual vehicles: steering wheel angle, acceleration/braking torques, etc.

\begin{figure*}[!ht]
\centering
\includegraphics[scale=0.6]{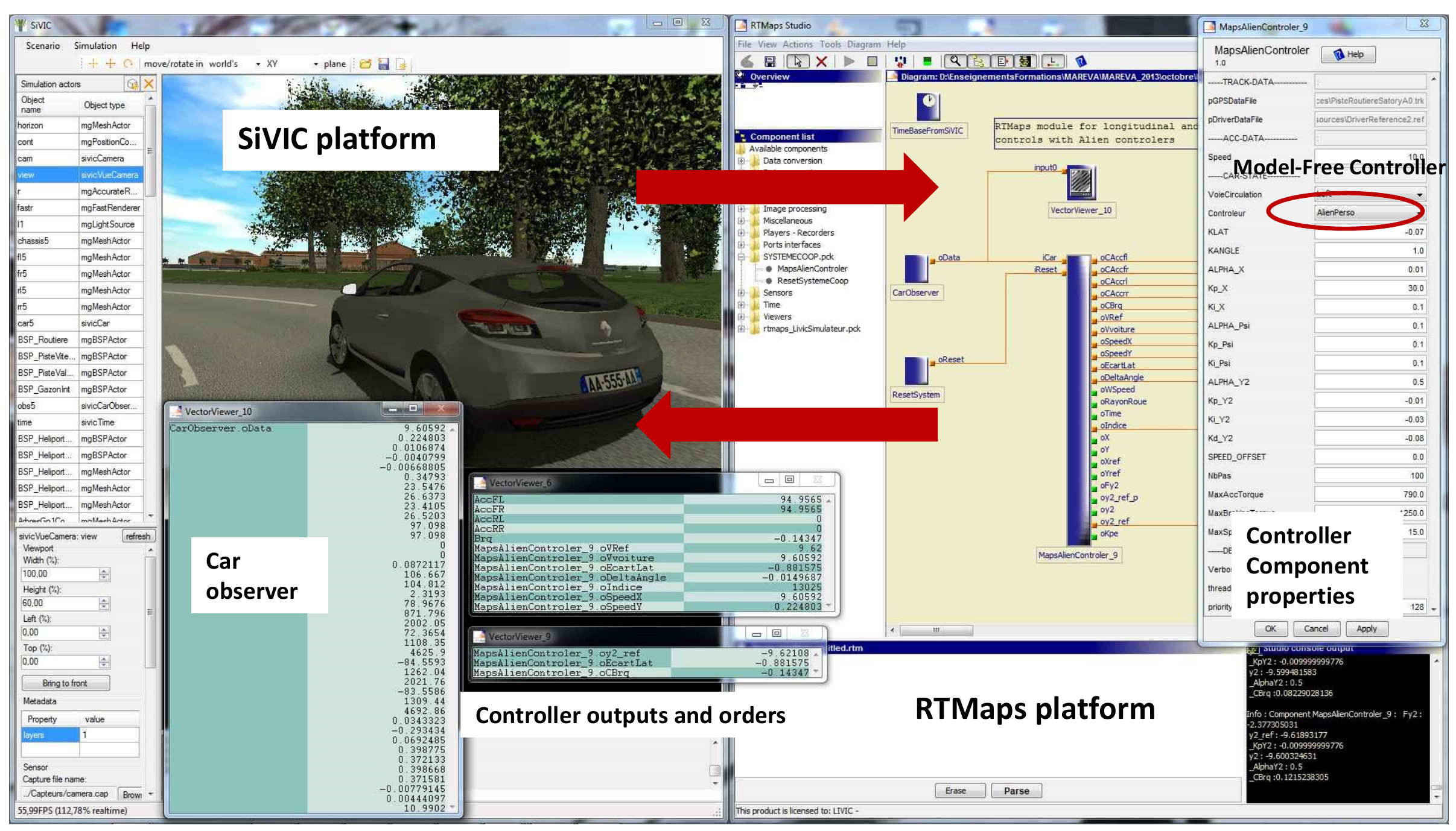}
\caption{Real-time implementation of the model-free control with interconnected platforms RTMAPS/pro-SiVIC}
\label{Inter_Sivic_RTMAPS}
\end{figure*}

\begin{figure*}[!ht]
\centering
\includegraphics[scale=0.65]{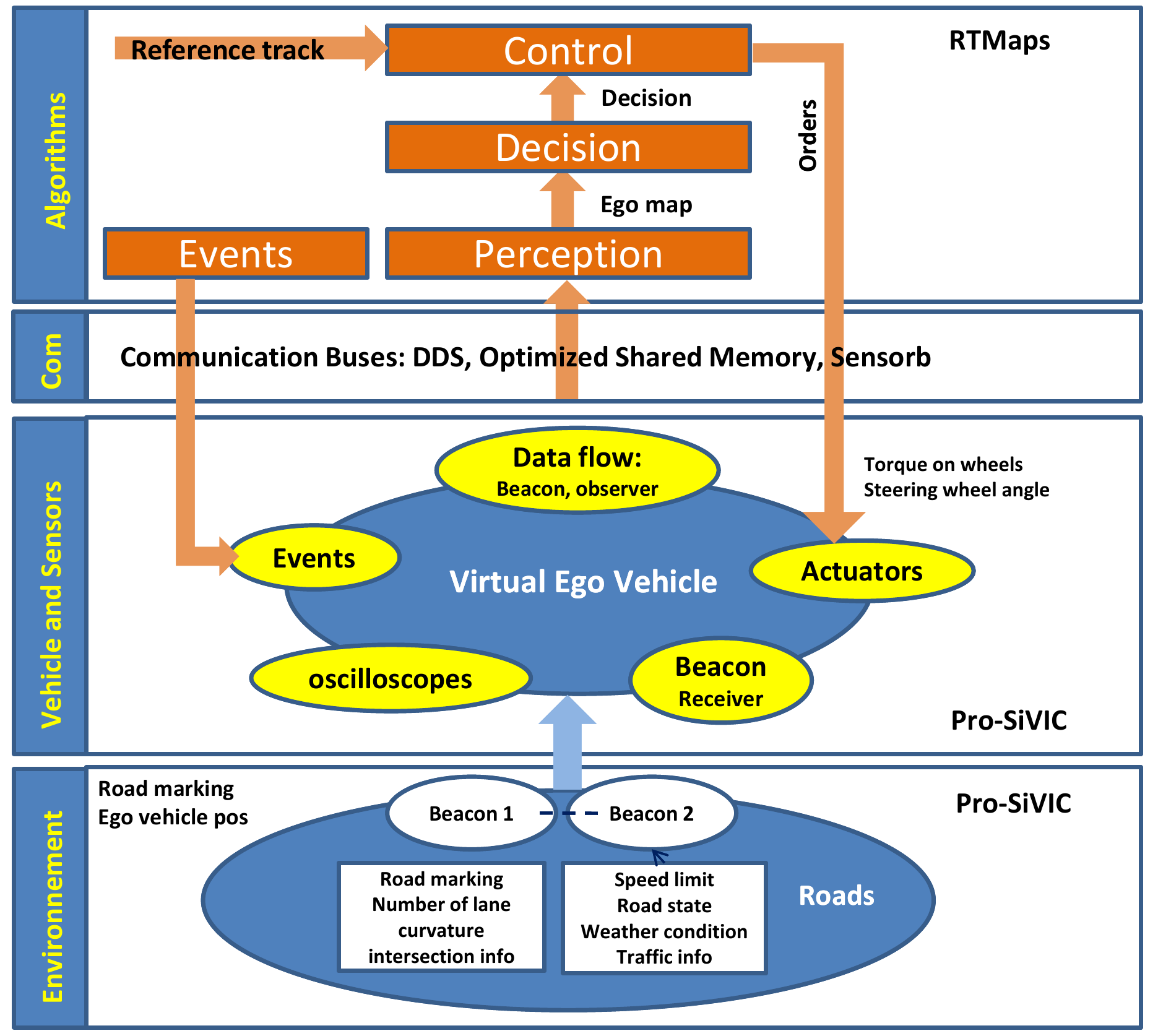}
\caption{General diagram of software in the loop model-free controller with functional layers}
\label{Inter_Sivic}
\end{figure*}

From the different modules and functionalities available in both pro-SiVIC and RTMaps platforms, and following the diagram block of Fig. \ref{diag_valid_control_NB_V1}, we have implemented a complete operational architecture in order to test and to evaluate the model-free controller with a real-time generated reference. In this architecture, shown in Fig. \ref{Inter_Sivic_RTMAPS} and Fig. \ref{Inter_Sivic}, pro-SiVIC provides the virtual environment, the complex vehicle dynamics modeling, and the sensors simulation. Then data coming from a vehicle observer sensor are sent towards RTMaps platform by using a dedicated interface library. This "`observer"' sensor, which is a state vector with 40 parameters, includes all informations concerning the ego-vehicle state and dynamics (positioning, angles, speeds, angle speeds, accelerations, wheel speeds, ...). In RTMaps, we take into account these vehicle observer data in order to calculate the longitudinal and lateral control outputs. Then, these outputs are sent back to pro-SiVIC's vehicle and more specifically to the virtual actuators. 
In addition, an events mechanism has been implemented in order to apply some dynamic constraints in the simulation (obstacle appearance, vehicle parameters modification, ...). In order to provide speed limit constraints in different areas, infrastructure sensors like road side beacons have been put on the virtual environment. The real-time software in the loop implementation with some data viewers is presented in Fig. \ref{Inter_Sivic_RTMAPS}.  

\subsubsection{Simulation results with the interconnected platforms pro-SiVIC and RTMaps}
\begin{figure*}[!ht]
\centering
\includegraphics[scale=0.4]{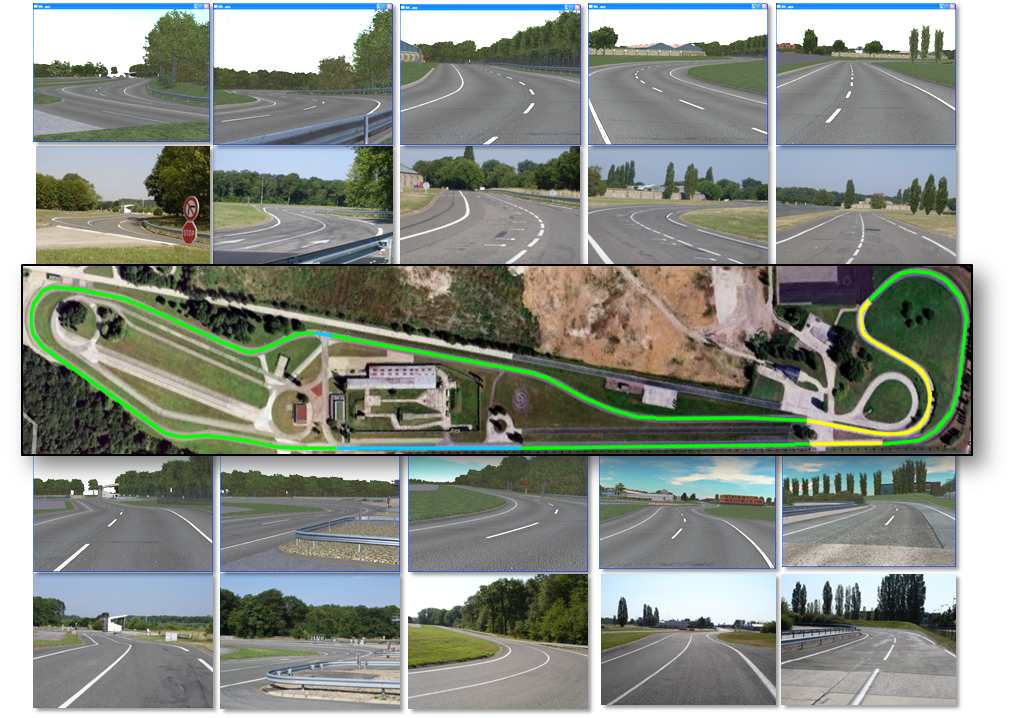}
\caption{Satory's test track (virtual, real, and bird view)}
\label{Satory_test_track}
\end{figure*}

In order to test the model-free control in realistic scenario, the Satory's test tracks have been used. The modeling of these tracks in the pro-SiVIC Platform was made from surveyor's data with a centimeter accuracy including road coordinates and road geometry. In addition, the entire virtual infrastructure corresponds to the real infrastructure present on the real tracks. These tracks are shared into three different areas. The first is the "routiere" track with a range of 3.4  km corresponding to a national road with very strong curvatures. The second track is a speedway (2 km) similar to a highway scene. The third track is the "`val d'or" area corresponding to a rural area. In our tests, we only used the first track. The level of reality between the track modeling in the simulation platform and the actual track can be seen in Fig. \ref{Satory_test_track}. Moreover, the inputs/outputs used in this second test are the same as those used in the first one (see Table \ref{Inputs_outputs}).

\begin{figure}[!ht]
\centering
\includegraphics[scale=0.57]{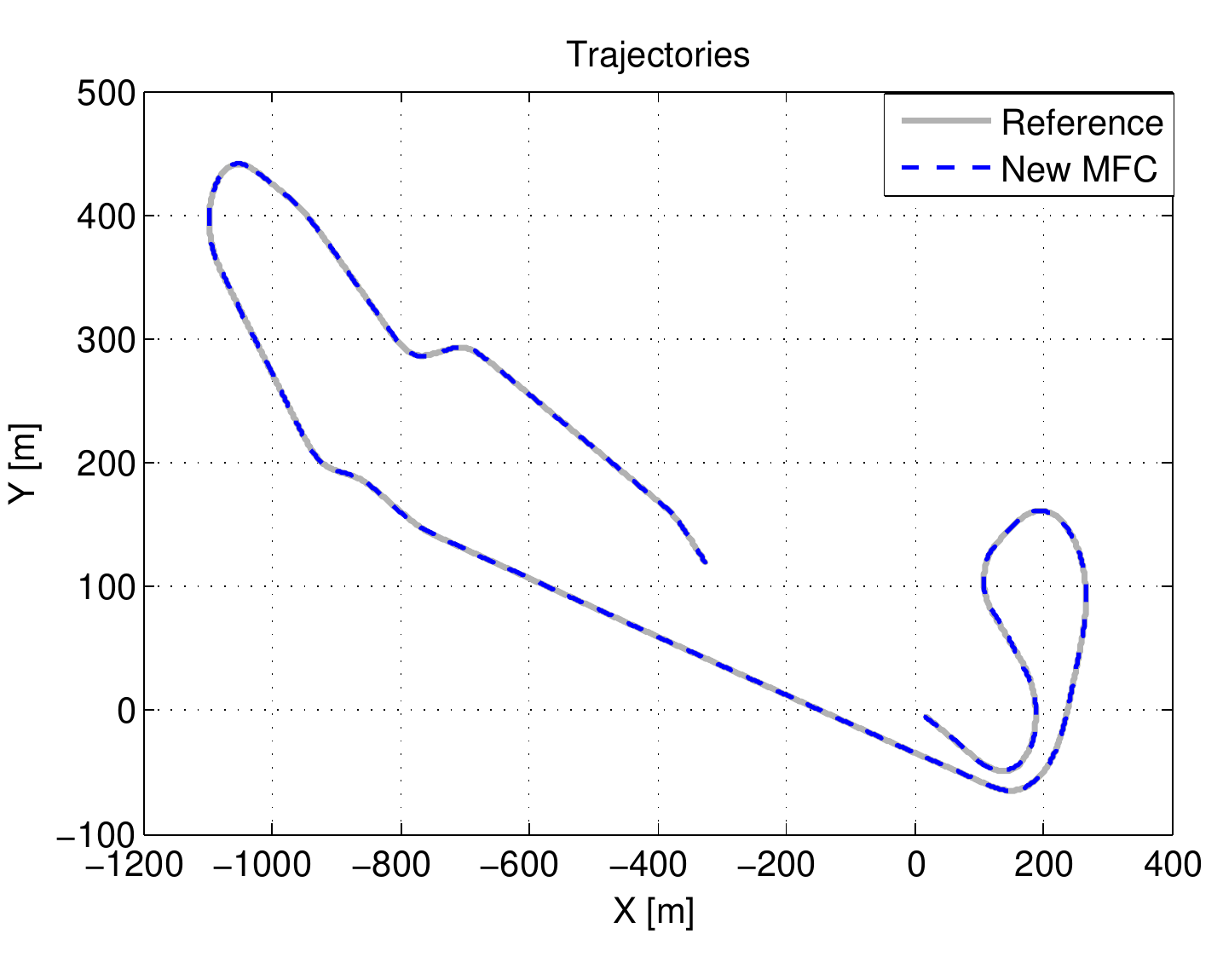}
\caption{Reference trajectory versus the simulated closed-loop trajectory}
\label{X_Y_Sivic}
\end{figure}

\begin{figure}[!ht]
\centering
\includegraphics[scale=0.57]{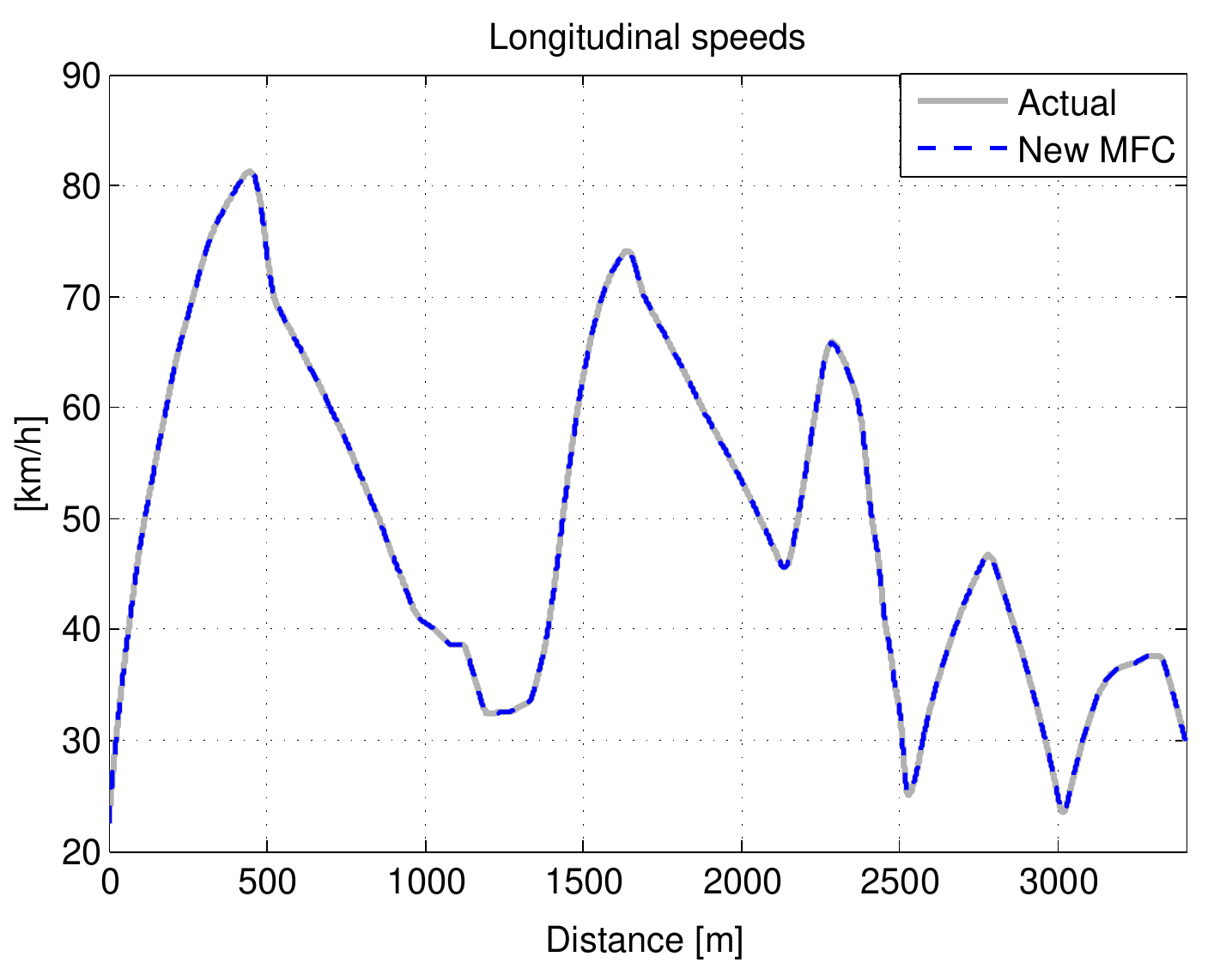}
\caption{Longitudinal speed: actual and simulated}
\label{Vx_SiVIC}
\end{figure}

\begin{figure}[!ht]
\centering
\includegraphics[scale=0.57]{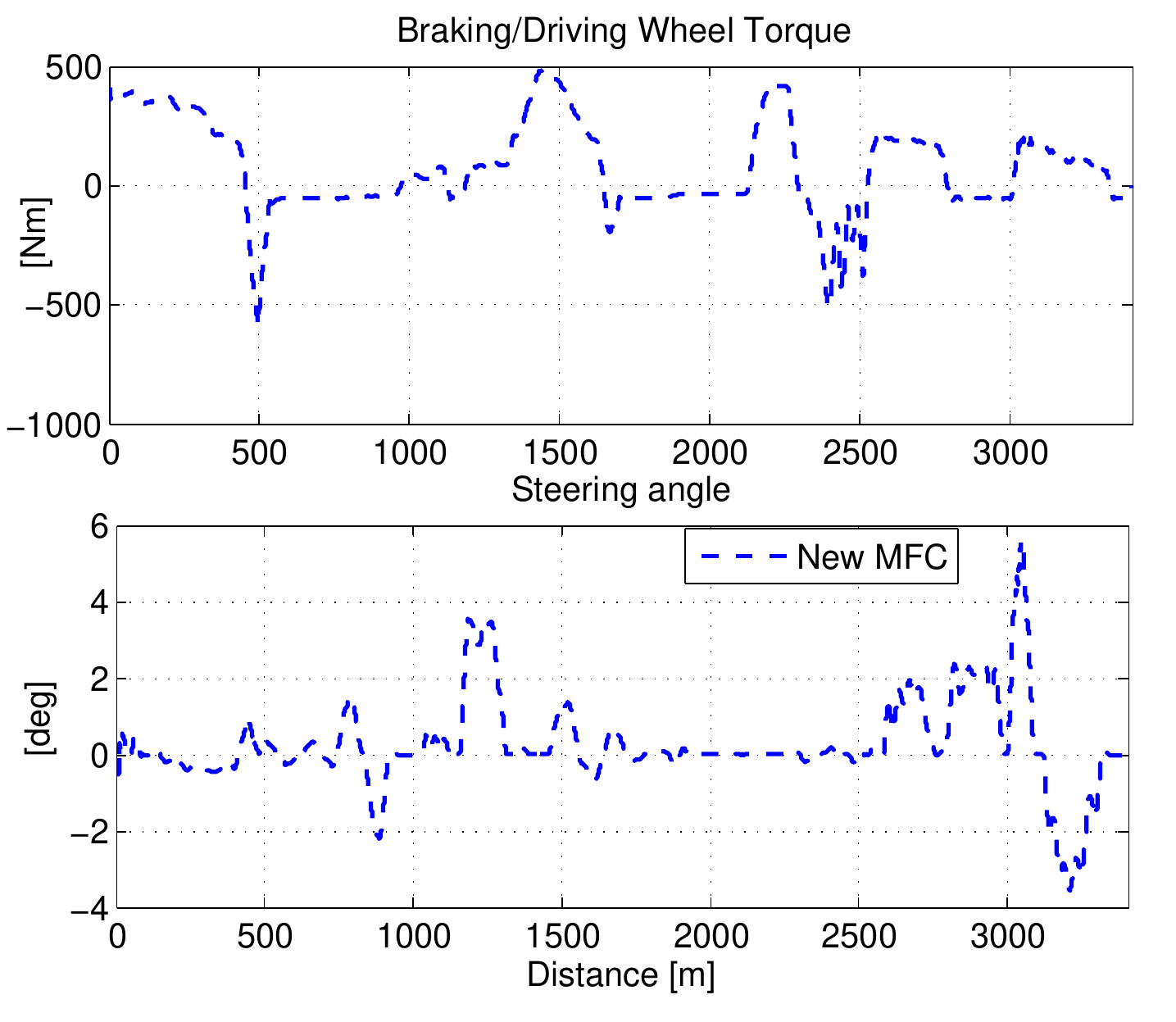}
\caption{Control inputs: Wheel torques and steering angles control signals}
\label{Tw_Delta_SiVIC}
\end{figure}

\begin{figure}[!ht]
\centering
\includegraphics[scale=0.57]{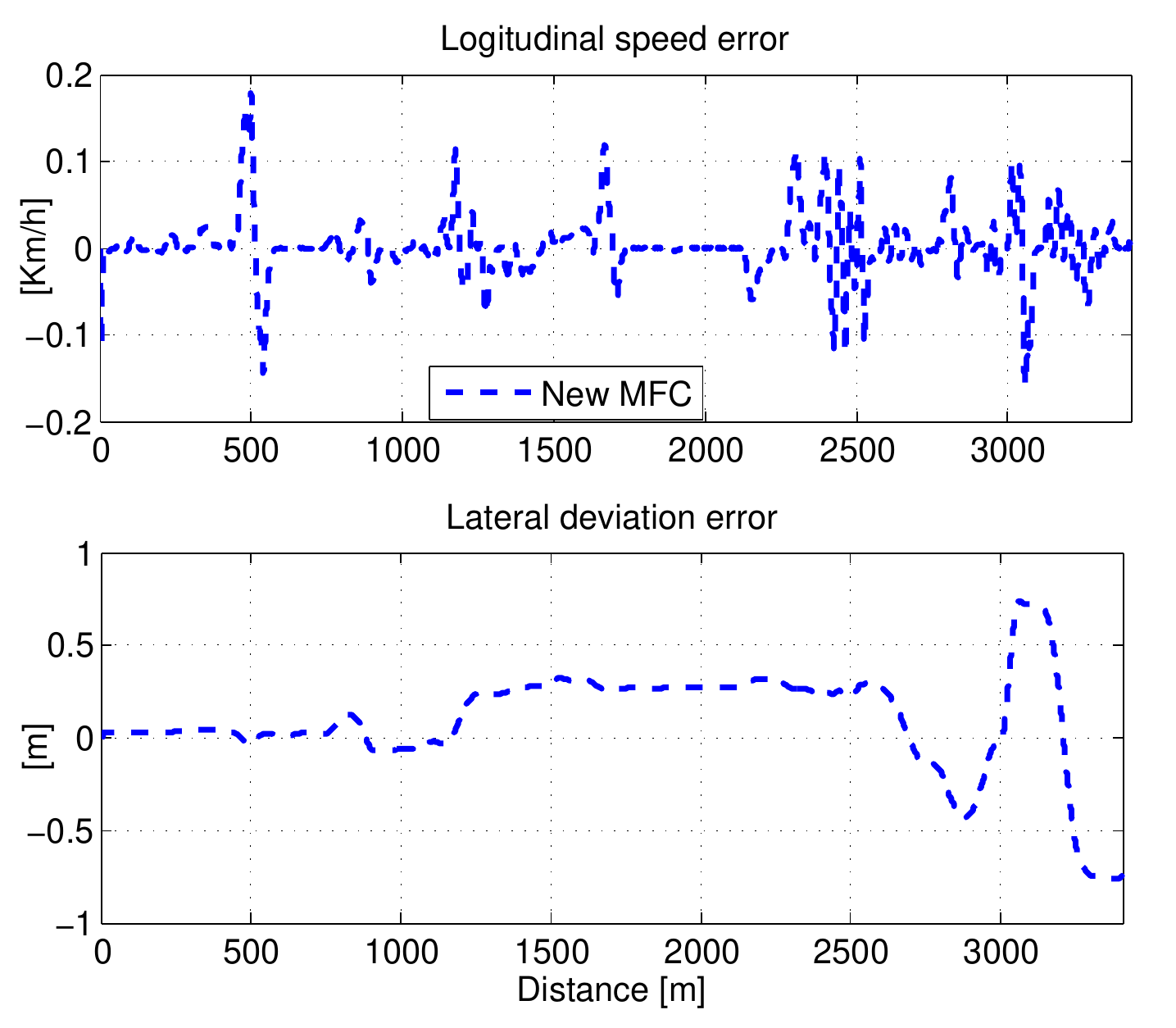}
\caption{Tracking errors on longitudinal speed and lateral deviation}
\label{EVx_Ey_SiVIC}
\end{figure}

Using this track, the Figs. \ref{X_Y_Sivic}, \ref{Vx_SiVIC}, \ref{Tw_Delta_SiVIC} and \ref{EVx_Ey_SiVIC} give the simulation results obtained by implementing the control law under the interconnected pro-SiVIC/RTMaps platforms. These results confirm the efficiency of the proposed control law even under a complex and a full simulation environment very close to real conditions. The tracking performances, in terms of longitudinal speed and lateral deviation tracking errors, are also depicted in Fig. \ref{EVx_Ey_SiVIC}. With this scenario included very strong curvatures, we can observe a little degradation of the lateral deviation accuracy. Nevertheless the results stay in an acceptable domain allowing to control position of the vehicle in the current lane. About the speed profile, the vehicle follows closely the reference with an absolute error lower than 0.2 ~km/h. 

\begin{figure}[!ht]
\centering
\includegraphics[scale=0.55]{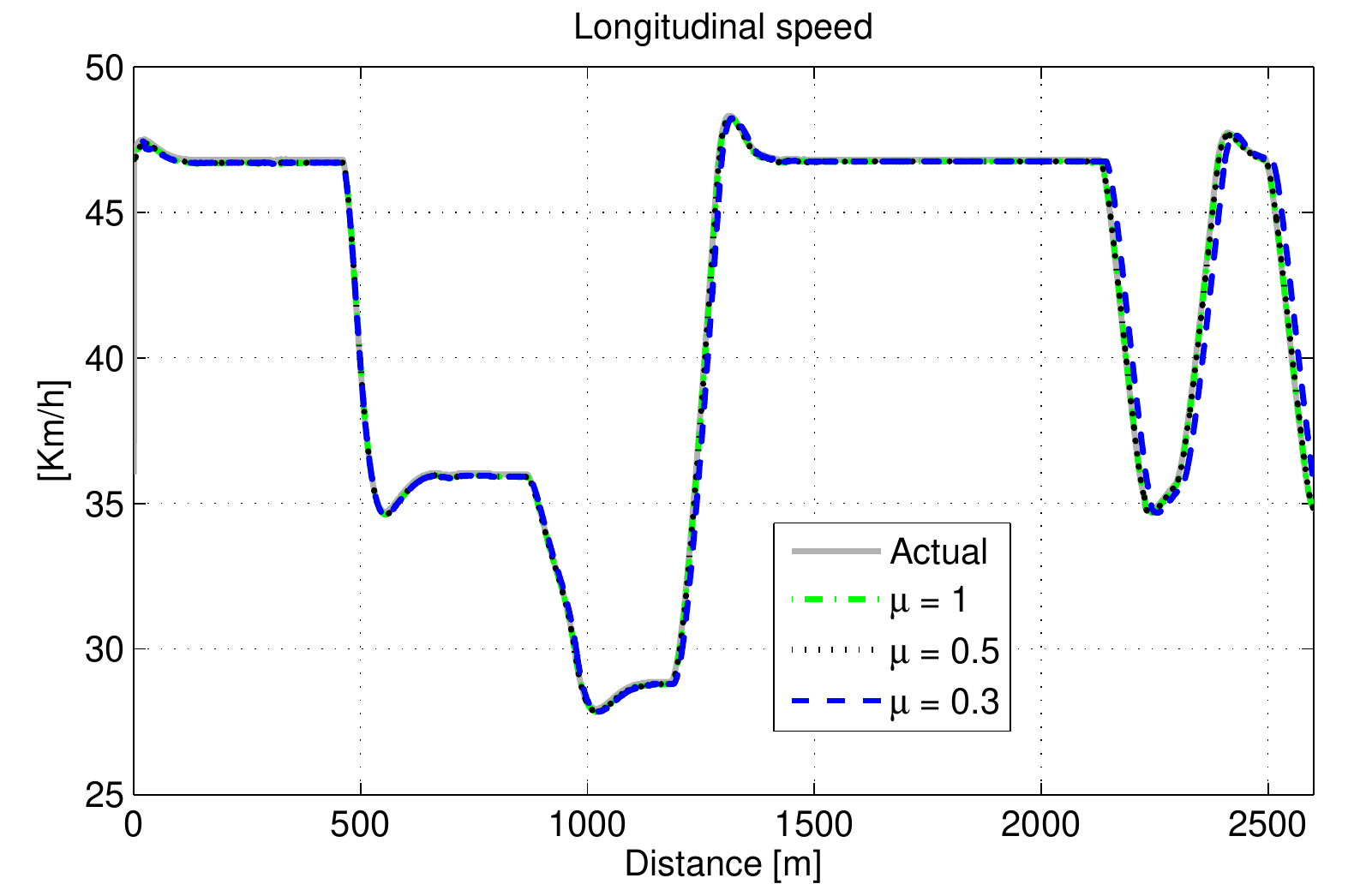}
\caption{Longitudinal speed: actual and simulated for $\mu = 1$, $\mu = 0.5$ and $\mu = 0.3$}
\label{Vx_mu1_05_03}
\end{figure}

\begin{figure}[!ht]
\centering
\includegraphics[scale=0.55]{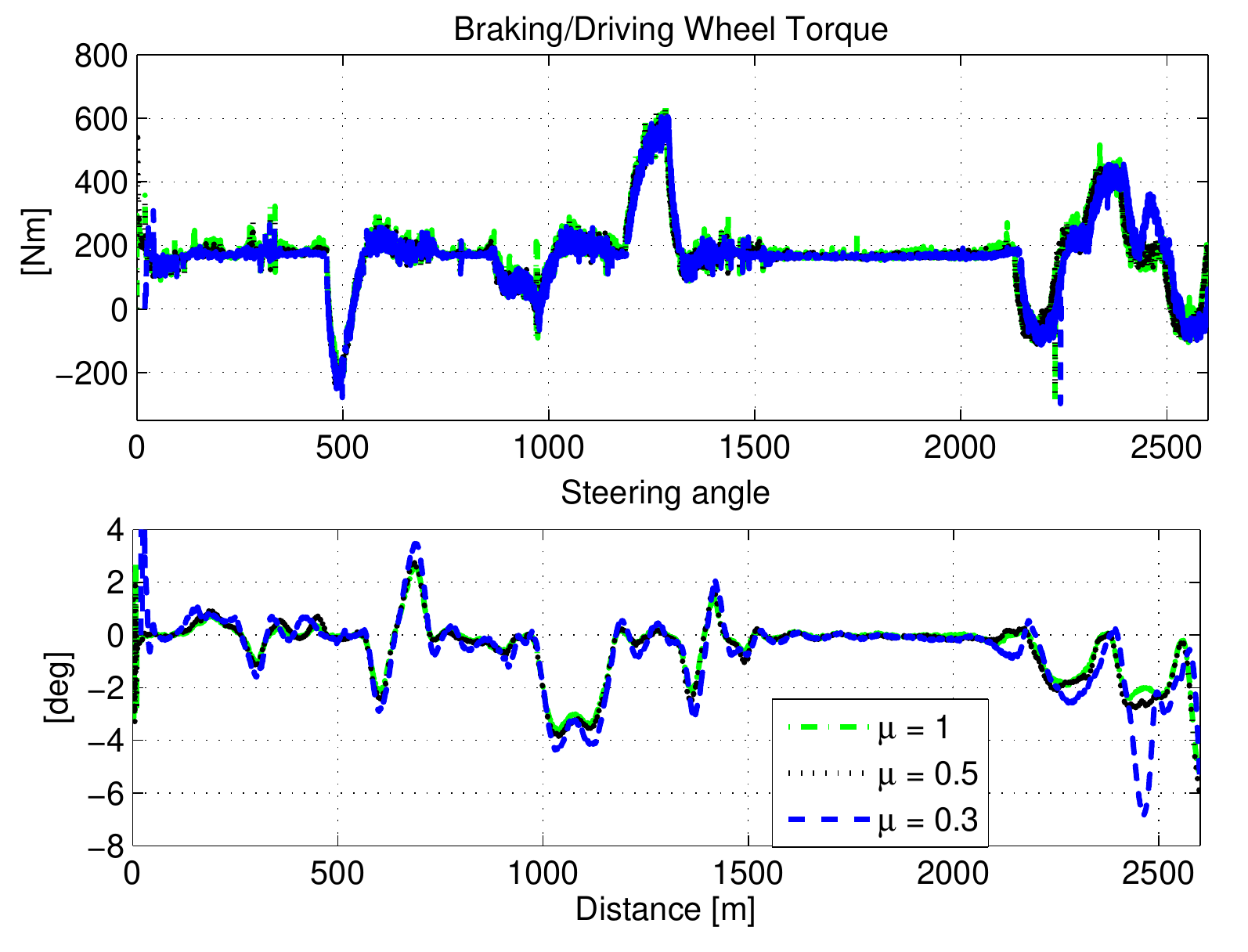}
\caption{Control inputs: Wheel torques and steering angles control signals  for $\mu = 1$, $\mu = 0.5$ and $\mu = 0.3$}
\label{Cw_delat_mu_1_05_03}
\end{figure}

\begin{figure}[!ht]
\centering
\includegraphics[scale=0.55]{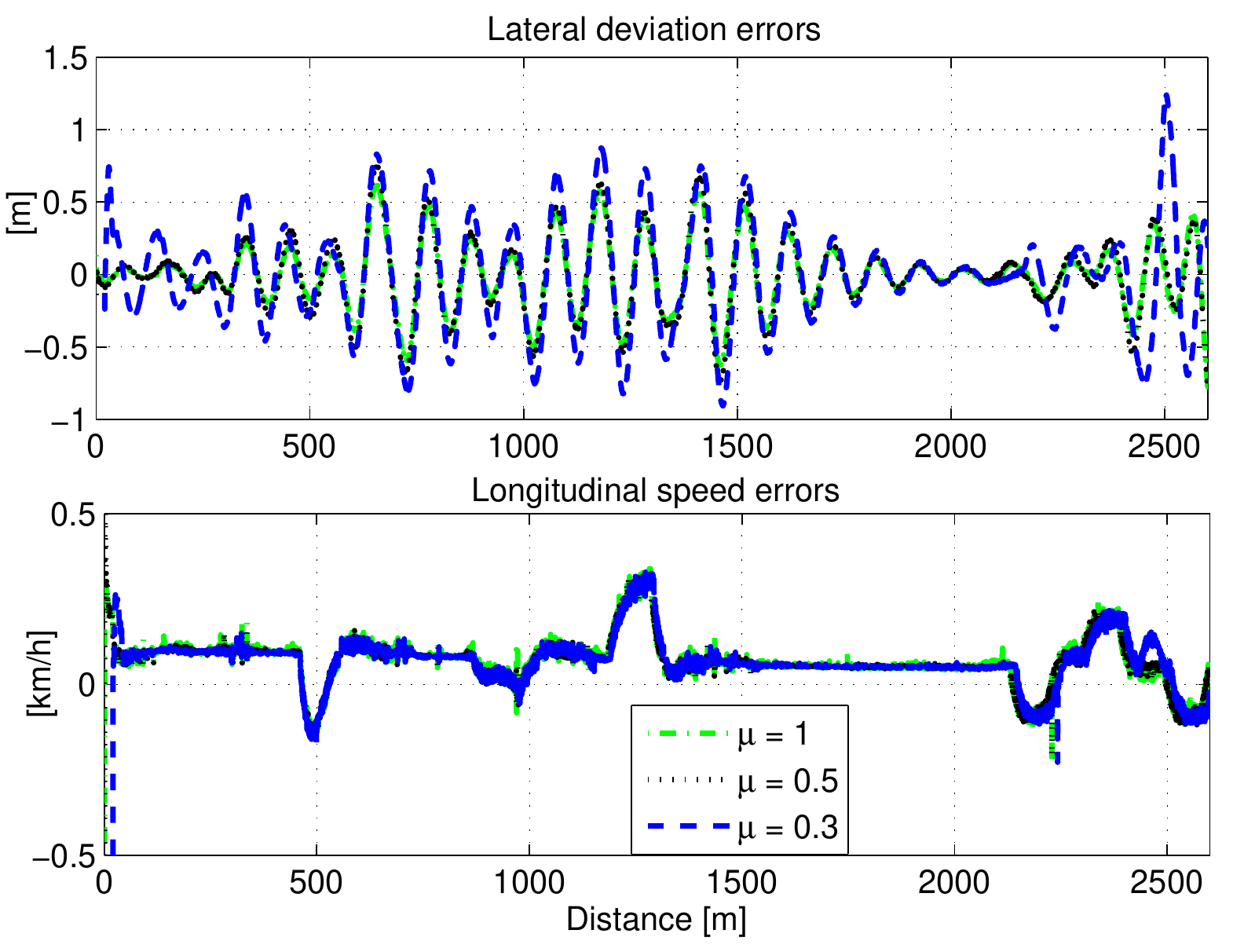}
\caption{Tracking errors on longitudinal speed and lateral deviation  for $\mu = 1$, $\mu = 0.5$ and $\mu = 0.3$}
\label{EVx_Ey_mu1_05_03}
\end{figure}

Moreover, Figs. \ref{Vx_mu1_05_03}, \ref{Cw_delat_mu_1_05_03} and \ref{EVx_Ey_mu1_05_03} show three simulation tests with three values of the road friction coefficient. We can observe that these tests highlight the performance of the model-free control approch under low road friction coefficient and high longitudinal speed.

\section{Conclusion}
\label{Section_5}

In recent years, many researches and many studies are carried out to allow the development of co-pilot functions for autonomous driving. 
In order to converge to a reliable, robust and functional solution in driving limit conditions, it is essential to be able to find answers to many research problems affecting both the perception of the environment, interpretation of road scenes, the safe path planning, the decision making and finally the control/command of the vehicle. In our work, we mainly address the final stage to produce control input for the vehicle control. Generally, in this stage, solutions require knowledge of vehicle dynamics and a accurate estimation of both the parameters and the variables of these dynamics models (often complex). Moreover, in real conditions, obtaining knowledge on these complex dynamic models is, in general, difficult and needs to implement expensive and complex means (sensors, hardware architecture). It is the same problem for the ground truth generation. Nevertheless, in case of poor estimation of the attributes of these dynamic models, the control algorithms could produce orders that will lead the system to diverge and by extension to dangerous maneuvers for the vehicle. In this paper, we propose a new approach of vehicle control using no vehicle evolution model. To demonstrate and to prove the relevance of such an approach, we have compared it with the results of several other more traditional approaches such as a PID controller and a flatness based controller. In order to test, to evaluate, and to validate this new approach, two simulation stages with different levels of dynamics vehicle modeling are used. 

The first level uses Matlab with a dynamic model of a Peugeot 406 (with 10 DoF). In this first stage of evaluation, the reference is provided by a trajectory obtained from a real Peugeot 406. The results clearly show a significant gain by using the model-free setting comparatively to the other. The accuracy obtained from the model-free controller is, in this first case of simulation, good enough in order to use it in an autonomous driving application. 

Another conclusion can also be inferred from the first results. Indeed, the use of the model-free controller also shows that the question of vehicle control can be effectively addressed and processed without need to a complex dynamics vehicle modeling. Only simple methods and functions are needed in order to solve this problem.

Secondly, to validate this approach in a simulation environment very close to the prototyping conditions and real condition use, we have implemented the model-free approach in a complex platform made up of pro-SiVIC (simulation of vehicles, infrastructure and sensors) and RTMaps (management and processing of data flows. Also used in the real vehicle prototypes of LIVIC and CAOR laboratories). The interconnection of these two platforms has enabled to integrate the model-free controller in a SIL architecture. In addition to generating information from embedded sensors, pro-SiVIC has made it possible to generate a very accurate ground truth. Again, the results were very successful and have validated the relevance of this new model-free controller approach even in extreme driving conditions with very low curvature radii. For this second level of simulation, the vehicle was evolving on a very realistic modeling of Satory's test tracks.

Currently, the tests were carried out using sensor data without noise or malfunctions. In future works, we will discuss the impact of noise and failures of the embedded sensors on the robustness of this model-free approach (see \cite{ijc13}). With these new studies, we hope that we will prove that the model-free controller can remain sufficiently robust and sensors failure tolerant to ensure safe control stage enough longer in time to give back the vehicle control to the driver. This step is crucial to enable the deployment of safe driving automation applications.

\appendices

\section{Flatness-based and PID controllers}
\label{appendix1}

\subsection{A flatness-based controller}
After some straightforward computations, model \eqref{Non_linear_bicycle_model} can be rewritten in the following standard form (see \cite{Menhour13a} for details):

\begin{equation}
\label{affine_NL_modele}
\dot{x}=f(x,t)+g(x,t)u
\end{equation}

Moreover the flatness property \cite{Fliess95,levine,hsr} of system \eqref{affine_NL_modele} is established according to the following proposition.

\begin{proposition}
The following outputs: 

\begin{equation}
\label{flatness_outputs}
\left \{\begin{array}{l}
z_1=V_x \\ [1mm]
z_2= L_f m V_y -I_z \dot{\psi}
\end{array}
\right.
\end{equation}
are flat outputs for system \eqref{affine_NL_modele}. 
\end{proposition}

\begin{IEEEproof}
Some algebraic manipulations (see \cite{Menhour13a} for more details) yield:
\begin{equation}
\label{x_A_y1_y2_y2p}
\begin{array}{c}
x=
\left[
\begin{array}{ccc}
V_x & V_y & \dot{\psi}
\end{array}
\right]^T
\begin{array}{c}
=\end{array}\\
\left[
\begin{array}{c}
z_1\\ [2mm]
\frac{z_2}{L_f m} -  \frac{I_z }{L_f m} \left(\frac{L_f m z_1\dot{z}_2 + C_r(L_f+L_r)z_2 }{C_r(L_f+L_r)(I_z -L_rL_fm)+ (L_f mz_1)^2 }\right)
\\ [2mm]
-\left(\frac{L_f m y_1\dot{z}_2 + C_r(L_f+L_r)z_2 }{C_r(L_f+L_r)(I_z -L_rL_fm)+ (L_f mz_1)^2 }\right)
\end{array}
\right]
\end{array}
\end{equation}

and

\begin{equation}
\label{U_B_y1_y2_y2p}
\begin{array}{c}
\left[
\begin{array}{c}
\dot{z}_1\\
\ddot{z}_2
\end{array}
\right]
= \Delta(z_1,z_2,\dot{z}_2) \left(
\begin{array}{c}
u_1\\
u_2
\end{array}
\right)+\Phi(z_1,z_2,\dot{z}_2) 
\end{array}
\end{equation}

The terms $\Delta_{11}$, $\Delta_{12}$, $\Delta_{21}$, $\Delta_{22}$, $\Phi_1$ and $\Phi_2$ of the matrices $\Delta$ and $\Phi$ are in detailed in \cite{Menhour13a}. Thus

\begin{equation}
\label{u_B_y1_y1p_y2_y2p_y2pp}
\begin{array}{c}
u=\left[
\begin{array}{c}
T_\omega\\
\delta
\end{array}
\right]=
\Delta^{-1}(z_1,z_2,\dot{z}_2)
\left(
\left[
\begin{array}{c}
\dot{z}_1\\
\ddot{z}_2
\end{array}
\right]
- \Phi(z_1,z_2,\dot{z}_2)
\right)
\end{array}
\end{equation}
with $r_x=1$ and $r_u=2$. Consequently, the system \eqref{affine_NL_modele} is flat with flat outputs \eqref{flatness_outputs}. Then, in order to track the desired output $z_1^{ref}$ and $z_2^{ref}$, set

    \begin{equation}
\label{lin_contro}
  \left [
  \begin{array}{c}
    \dot{z}_1 \\
    \ddot{z}_2
  \end{array} \right]=
  \left [
  \begin{array}{c}
  \dot{z}_1^{ref}+K_1^1e_{z_1}+K_1^2\int{e_{z_1} dt}\\[2mm]
\ddot{z}_2^{ref}+K_2^1\dot{e}_{z_2}+K_2^2e_{z_2}+K_2^3\int{e_{z_2} dt}
  \end{array} \right]
    \end{equation}
where, $e_{z_1}=z_1^{ref}-z_1=V_x^{ref}-V_x$ and
$e_{z_2}=z_2^{ref}-z_2$. Choosing the gains  $K_1^1$,
$K_1^2$, $K_2^1$, $K_2^2$ and $K_2^3$ is straightforward.

\end{IEEEproof}

The control law contains derivatives of measured signals, which are of course noisy. This estimation is performed using the recent advances algebraic estimation techniques in \cite{Fliess08, Mboup09,sira1,sira2}. 
\begin{remark}
\label{Remark_0}
Let us notice that the second flat output $z_2$ is the angular momentum of a point located on the axis between the centers of the rear and front axles.
\end{remark}

\subsection{PID controllers}
\label{Remark_1}

It seems interesting, in order to test the validity and the limits of the proposed control methods, to compare them with the classical well known PID controllers. 
Numerous methods to design PID controllers exist in the literature, see e.g. \cite{Astrom88, Ho98, Saeki06, Wang07}. For our case study, the gains $(K_p^{V_x},\, K_d^{V_x},\,K_i^{V_x})$ (gains of longitudinal PID$_{V_x}$ control) and $(K_p^{y},\, K_d^{y},\,K_i^{y})$ (gains of lateral deviation PID$_{y}$ control) are computed in vicinity of $V= 20 m/s$. Then, using the oscillation method \cite{Astrom88}, the chosen numerical values of the gains are $(K_p^{V_x} = 1.51,\, K_d^{V_x} = 0.52,\,K_i^{V_x} = 0.75)$ for PID$_{V_x}$ and $(K_p^{y} = 0.95,\, K_d^{y}=0.36,\,K_i^{y} = 46)$ for PID$_{y}$. Moreover, in order to avoid the impact of the noisy measurements on the longitudinal speed and the lateral deviation, a first order filter is associated to the derivative actions of these controllers (see \cite{Astrom88} for more details). The following PID controllers are used:

\begin{equation}
\label{PID_controllers}
\left \{ \begin{array}{lcl}
u_1(t) =  K_p^{V_x}e_{V_x}+ K_d^{V_x}\dot{e}_{V_x} + K_i^{V_x} \int e_{V_x} dt\\
u_2(t) =K_p^{y}e_{y}+ K_d^{y}\dot{e}_{y} + K_i^{y} \int e_{y} dt
\end{array}
\right.
\end{equation}

\begin{IEEEbiographynophoto}{Lghani MENHOUR}
obtained his Ph.D. in automatic control in 2010 from the Laboratoire Heuristique et diagnostic des syst\`emes complexes of the Universit\'e de Technologie de Compi\`egne, Compi\`egne, France. Since 2011, he is an Assistant Professor at the Laboratoire CReSTIC of the Universit\'e de Reims-Champagne-Ardenne, Reims, France. His research interests include linear and nonlinear systems, control and estimation, switched systems, Model-free control, algebraic approaches, intelligent transportation systems.
\end{IEEEbiographynophoto}

\begin{IEEEbiographynophoto}{Brigitte d'ANDR\'EA-NOVEL}
graduated from \'{E}cole Supérieure d'Informatique, \'{E}lectronique, Automatique in 1984. She received the Ph. D. degree from \'{E}cole Nationale Supérieure des Mines de Paris in 1987 and the Habilitation degree from Université Paris-Sud in 1995. She is currently a Professor of Systems Control Theory and responsible for the research group in Advanced Control Systems at the Center for Robotics from MINES ParisTech. Her current research interests include nonlinear control theory and applications to underactuated mechanical systems, control of wheeled vehicles with applications to automated highways. In that context she has developed regular collaborations with PSA-Peugeot-Citro\"{e}n and Valeo on ``Global Chassis Control'', ``Stop and Go algorithms'' and ``Automatic parking''. Moreover, she has also been interested in modeling and boundary control of dynamical systems coupling ODEs and PDEs, with applications to the control of irrigation canals and wind musical instruments. 
\end{IEEEbiographynophoto}

\begin{IEEEbiographynophoto}{Michel FLIESS}
works at the \'Ecole polytechnique (Palaiseau, France). In 1991 he invented with J. L\'evine, P. Martin, and P. Rouchon (Ecole des Mines de Paris, France), the notion of differentially flat systems which is playing a crucial role in many concrete situations all over the world. In 2002 he introduced with H. Sira-Ram\'{i}rez (CINVESTAV, Mexico City) algebraic estimation and identification techniques, which are most useful in control and signal processing. Ten years ago he started with C. Join (Universit\'e de Lorraine, Nancy, France) a model-free control setting which is now being used in a number of concrete industrial applications. He got several prizes, including three ones from the French Academy of Sciences. \end{IEEEbiographynophoto}

\begin{IEEEbiographynophoto}{Dominique GRUYER}
was born in France, in 1969. He received the M.S. and Ph.D. degree respectively in 1995 and 1999 from the University of Technology of Compiègne. Since 2001, he is a researcher at INRETS, into the perception team of the LIVIC department (Laboratory on interactions between vehicles, Infrastructure and drivers) and he works on the study and the development of multi-sensors/sources association, combination and fusion. His works enter into the conception of on-board driving assistance systems and more precisely on the carry out of multi-obstacle detection and tracking, extended perception, accurate ego-localization. He is involved for multi-sensor fusion tasks and sensors modeling and simulation in several European and French projects dealing with intelligent vehicles (HAVEit, Isi-PADAS, CVIS, CARSENSE, eMOTIVE, LOVe, ARCOS, MICADO, ABV, eFuture, CooPerCom, SINETIC, ...). He is responsible and the main inventor of the SiVIC platform (Simulation for Vehicle, Infrastructure and sensors). Since 2010, He is the team leader of the LIVIC's Perception team. For 5 years, he is a network researcher in AUTO21 (Canada). He was the co-founder and technical Director of the CIVITEC Company until 2015. Since April 2015, CIVITEC is a subsidiary of the ESI group company.
He is now the head of the LIVIC laboratory (since January 2015), a Research Director (since 2014) in IFSTTAR, and Perception system and data fusion Scientific Director (since April 2015) for ESI group (CIVITEC).
\end{IEEEbiographynophoto}

\begin{IEEEbiographynophoto}{Hugues MOUNIER}
obtained his Ph.D. in automatic control in 1995 from the Laboratoire des Signaux et Syst\`emes of the Universit\'e Paris Sud, Orsay, France. From 1998 to 2010 he has been with the Institut d'Electronique Fondamentale of the same University. He is currently Professor at the Laboratoire des Signaux et Syst\`emes. His research interests include automotive and real-time control, neuroscience, delay systems and systems modelled by partial differential equations
\end{IEEEbiographynophoto}

\end{document}